\documentclass[11pt, a4]{article}
\usepackage{amssymb,amsmath,latexsym,graphicx, bm, mathrsfs}
\usepackage[hidelinks,citecolor=blue,urlcolor=blue,colorlinks=true,bookmarks=false,hypertexnames=true]{hyperref} 
\usepackage{natbib} 
\setcitestyle{numbers,square} 
\usepackage{mathtools}
\usepackage{graphics,graphicx, cancel}
\usepackage[dvipsnames]{xcolor}
\usepackage{graphicx}
\graphicspath{ {./images/} }
\usepackage[british]{datetime2}
\usepackage[page,toc,titletoc,title]{appendix}
\usepackage{tocloft}
\usepackage{blindtext}
\usepackage{dsfont}
\usepackage{dirtytalk}

\usepackage{comment}
\usepackage{caption}
\usepackage{enumitem}
\usepackage{mathrsfs}
\usepackage{alphalph} 
\usepackage{changepage}
\usepackage{braket}
\newcounter{mainthm}

\newcounter{subthm}[mainthm]

\newtheorem{subtheoreminner}{Theorem}[subthm]

\begin{document}

\title{Kinematics of Acceleration-Induced Excitations\\in Confined Quantum Fields}
\vspace{-0.25cm}
\author{
Hemansh Shah$^{1}$, Sanved Kolekar$^{2}$
\\ \vspace{-0.5cm} \\
\small $^{1}$\textit{Indian Institute of Science, CV Raman Road, Bengaluru 560012, India} \\
\small $^{2}$\textit{Indian Institute of Astrophysics, Block 2, 100 Feet Road, Koramangala, Bengaluru 560034, India} \\
\texttt{\small Email: \href{mailto:hemansha@iisc.ac.in}{hemansha@iisc.ac.in}, \href{mailto:sanved.kolekar@iiap.res.in}{sanved.kolekar@iiap.res.in}}
}

\maketitle

\begin{abstract}
We investigate quantum field excitations in a rigid cavity that undergoes a transition from inertial motion to uniform acceleration while maintaining constant proper length. By constructing exact Bogoliubov transformations between inertial and accelerated mode bases, we analyze the induced excitations and identify a universal power-law decay in the excitation power spectrum and an alternating pattern in particle production. The spectrum’s dependence solely on the acceleration, and not on the size of the box or the observer’s position, highlights a kinematic universality akin to that seen in horizon thermodynamics. Generalization to time-dependent accelerations reveals a convolution structure for the Bogoliubov coefficients, with resonant oscillations selectively enhancing mode excitations. These results provide new analytical insights into the interplay between acceleration, confinement, and quantum excitations. 
\end{abstract}

\section{Introduction}

Quantum field theory in curved spacetime (QFTCS) provides a consistent framework for studying quantum fields in the presence of classical gravitational or non-inertial backgrounds. In contrast to quantum gravity, the spacetime metric in QFTCS is treated as a fixed, classical entity while the matter fields are quantized. This semiclassical approach has yielded some of the most profound results in theoretical physics, including Hawking radiation~\cite{hawking1974black} and the Unruh effect~\cite{unruh1976notes}, both of which demonstrate that the concept of vacuum is observer-dependent. These results have fundamentally altered our understanding of quantum fields and their interaction with spacetime geometry, with implications ranging from black hole thermodynamics to the generation of cosmological perturbations~\cite{Parker1966,parker2017fiftyyearscosmologicalparticle}.

The Unruh effect predicts that an observer undergoing uniform acceleration perceives the Minkowski vacuum as a thermal bath, with a temperature proportional to the acceleration. This observer-dependent notion of particles reflects the non-invariance of vacuum states under general coordinate transformations. A closely related phenomenon is Hawking radiation, where particle creation arises near black hole horizons due to the causal structure of spacetime. Both effects exemplify how acceleration or curvature can mix positive and negative frequency modes, leading to observable particle creation.

Another closely related manifestation of vacuum fluctuations is the Casimir effect, first predicted by Casimir in 1948~\cite{casimir1948attraction}, where static boundaries alter the vacuum energy, resulting in a measurable force. Its time-dependent extension, the dynamical Casimir effect (DCE), involves particle production caused by non-stationary boundary conditions. Moore’s seminal work~\cite{moore1970quantum} demonstrated that a moving mirror in a one-dimensional cavity can convert vacuum fluctuations into real particles. While the Unruh effect attributes such excitations to non-inertial motion of the observer, the DCE attributes them to the dynamical evolution of the boundaries themselves.

The theoretical development of the DCE spans several decades. Initial analyses considered scalar fields in (1+1)-dimensional spacetime with a single accelerated mirror~\cite{DeWitt1975,DaviesFulling1977}, followed by studies of fields confined between two moving mirrors, particularly under periodic boundary motion~\cite{DodonovKlimov1992,DodonovKlimovNikonov1993}. The framework was later extended to higher-dimensional spacetimes~\cite{Milton2001}, to spinor and vector fields~\cite{Yablonovitch1989,Milton2004}, and to curved backgrounds~\cite{lock2017dynamical}. Further generalizations include partially reflective boundaries~\cite{Dalvit2011,BaezCamargo2024}, quantized boundary motion~\cite{Kulagin1996}, and stochastic trajectories~\cite{Mantinan2023}. This effect has since been extensively studied, both theoretically and experimentally.~\cite{dodonov2020fifty,dodonov2025dynamical}. These studies collectively establish the DCE as a robust phenomenon arising from the coupling between geometry and quantum vacuum fluctuations.

Several theoretical works have highlighted deep correspondences between acceleration-induced and boundary-induced particle creation. Nikishov and Ritus~\cite{NikishovRitus1995} showed that the spectrum of radiation emitted by an accelerated mirror in (1+1) dimensions is identical to that produced by an accelerating charge in (3+1) dimensions. Ritus further extended this correspondence to spinor fields~\cite{Ritus1996functionalidentity}. Mendonça \emph{et al.}~\cite{mendonca2008vacuum} demonstrated that photon creation inside a vibrating cavity is equivalent to time refraction and closely related to the Unruh effect in free space. These connections underscore the unifying role of acceleration and time-dependent boundary conditions in qu  antum field excitation.

On the experimental side, the DCE was first observed unambiguously by Wilson \emph{et al.}~\cite{Wilson2011} in a superconducting coplanar waveguide, and subsequently by Lähteenmäki \emph{et al.}~\cite{Lahteenmaki2013} in a Josephson metamaterial. More recent proposals include the use of atoms in accelerated cavities~\cite{lochan2020detecting}, circuit-QED analogs with tunable boundary conditions~\cite{corona2016dynamical}, hybrid optomechanical platforms~\cite{lan2024dynamical}, nanophotonic configurations~\cite{gangaraj2024dynamical}, and dispersion-engineered optical fibers~\cite{Vezzoli2018}. Notably, Vezzoli \emph{et al.} reported photon-pair creation from vacuum fluctuations through periodic modulation of the refractive index, establishing an optical analogue of the DCE. Numerical simulations of moving-mirror setups~\cite{olmedo2024numerical} further reinforce the theoretical and experimental feasibility of detecting such effects.

Most discussions of the DCE involve cavities whose physical length changes in time. However, several studies have examined particle creation in non-inertial cavities of fixed proper length, allowing direct comparison between inertial and accelerated frames without the additional complexity of boundary motion. Sorge~\cite{sorge2006dynamical} analyzed a cavity subjected to a short acceleration burst, while Levin~\cite{levin1992quantum} studied detector response in accelerated cavities. Boyer~\cite{boyer2012contrasting} contrasted classical zero-point radiation with quantum field theory predictions in non-inertial frames. Related analyses have explored entanglement degradation in non-inertial motion~\cite{Fuentes2005bosonic,alsing2006entanglement}, mode mixing in accelerated cavities~\cite{Bruschi2012Voyage,bruschi2013mode,Bruschi_2013_withoutparticlecreation}, and cosmological analogues~\cite{Ball2006expansion,martin2010cosmological}. In the limit of small cavity acceleration, the behavior of different types of quantum fields in a cavity with fixed dimensions has been explored in depth, with the development of a `small acceleration formalism'~\cite{Bruschi2012Voyage,Friis2012,Friis2013}. These studies emphasize the fundamental interplay between confinement, acceleration, and observer dependence in quantum field theory.

In this work, we consider a massless scalar field in (1+1)-dimensional Minkowski spacetime, confined within a box of constant proper length. The box is initially inertial and then undergoes a sudden transition to uniform acceleration while maintaining its proper size. This configuration enables a controlled analysis of acceleration-induced excitations without introducing geometric distortions due to changing boundary separation. The setup thus provides a direct bridge between the Unruh and dynamical Casimir frameworks.

We employ Bogoliubov transformations to relate the field modes before and after the transition. These transformations encapsulate the mixing between positive and negative frequency components and quantify particle creation. Similar methods have previously been employed to study entropy production~\cite{hasegawa2023relativistic}, information scrambling~\cite{hasegawa2021relativistic}, entanglement degradation~\cite{Bruschi2012Voyage}, and mode mixing in accelerated cavities~\cite{hasegawa2023entropy}. We extend these analyses to obtain exact and approximate expressions for the Bogoliubov coefficients relevant to our configuration.

The novelty of the present work lies in the exact treatment of a single transition from inertial to accelerated motion while preserving the cavity’s proper length. In contrast to previous approaches that consider arbitrary mirror trajectories within a single reference frame, we describe the field in coordinate systems naturally adapted to each regime. Where possible, we obtain closed-form expressions for the Bogoliubov coefficients, allowing a detailed characterization of the excitation spectrum and its asymptotic behavior.

Section 2 introduces the coordinate systems and describes the physical setup. We present the mode expansions for both inertial and accelerated regimes and impose continuity and differentiability conditions to match the fields across the transition. Section~3 derives integral expressions for the Bogoliubov coefficients, evaluates them exactly using special functions and computes the excitation spectrum as a sum over the squares of the Bogoliubov $\beta$-coefficients. Section~4 focuses on the low-acceleration limit ($aL \ll c^2$), deriving first-order expressions and analyzing mode dependence of the excitation spectrum. Section~5 evaluates the exact coefficients for large mode numbers at arbitrary acceleration, revealing a well-defined continuum limit. Section~6 generalizes the setup to arbitrary acceleration profiles, obtaining exact integral forms for the Bogoliubov coefficients and closed-form expressions to first order in acceleration. Finally, Section~7 discusses possible experimental realizations, including noise estimates, detection probabilities, and resonance effects in oscillating cavities. The implications are discussed in the last section.

Throughout, we use natural units $c = \hbar = 1$, except where explicit units are required. The results presented here provide analytical insight into acceleration-induced quantum excitations and help clarify the relationship between Unruh and Casimir-type phenomena within a unified framework.

\section{Field in a rigid box}

We consider a (1+1)-dimensional flat Minkowski spacetime in which a massless, real scalar field \( \phi(x,t) \) is confined within a rigid box of fixed proper length \( L \). The field obeys the massless Klein-Gordon equation and satisfies Dirichlet boundary conditions at the worldlines of the box boundaries, ensuring that the field amplitude vanishes at both ends. In the inertial phase, the field admits a discrete set of normal modes corresponding to standing waves confined within the box, and the Minkowski vacuum is defined as the state annihilated by the corresponding inertial mode annihilation operators.

At a particular instant, the box starts to uniformly accelerate while preserving its proper length, so that each point of the box undergoes a motion consistent with Born rigidity. The boundaries thus follow uniformly accelerated trajectories that maintain constant proper separation in Rindler coordinates. During the accelerated phase, the field modes are naturally defined with respect to the timelike Killing vector corresponding to continuous Lorentz boosts in the Rindler spacetime, leading to a distinct notion of positive frequency modes. The transition from inertial to uniformly accelerated motion therefore induces a nontrivial Bogoliubov transformation between the two complete sets of modes. Our aim is to analyze the resulting field excitations and the particle content, particularly the frequency dependence, perceived by an observer co-moving with the rigid box after the onset of acceleration. We begin by formally defining the quantities relevant for the analytical calculations.

\subsection{Setup}

\begin{figure}
    \centering
    \includegraphics[width=0.35\linewidth]{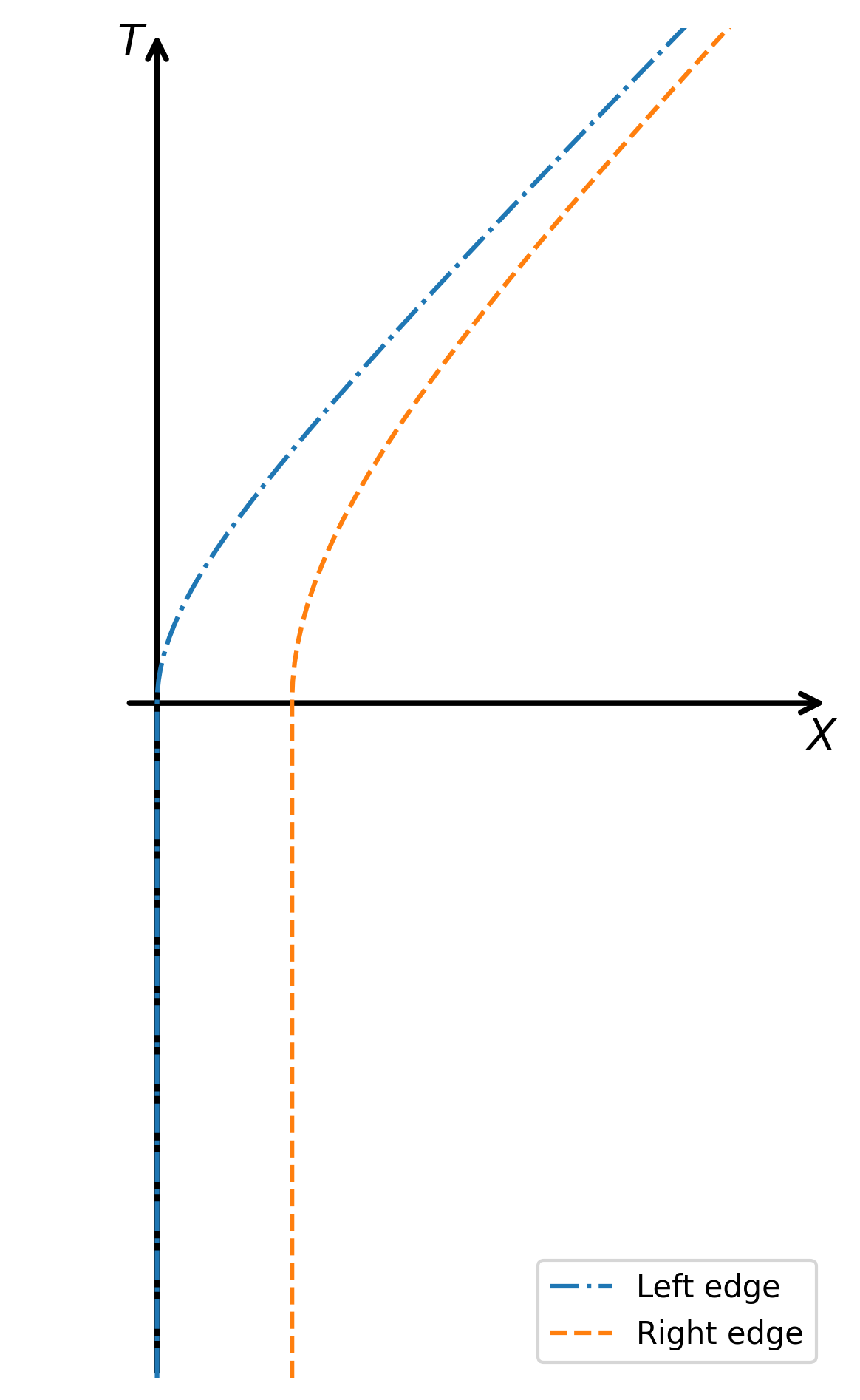}
    \caption{Worldlines of the left and right edges of the box confining the field, plotted in Minkowski coordinates \((T,X)\). The box transitions from inertial motion to uniform acceleration at \(T = 0\).}
    \label{fig:box}
\end{figure}

During the inertial phase, it is convenient to employ Minkowski coordinates \((T, X)\), in which the box remains at rest. We choose the left edge of the box to be located at \(X = 0\), and define \(T = 0\) as the instant at which the box begins to accelerate. The worldlines of the box edges in Minkowski coordinates are illustrated in Fig.~\ref{fig:box}.

Once the box begins to accelerate, it is natural to switch to Rindler coordinates \((t, x)\), which are adapted to uniformly accelerated observers. We define \(t = 0\) to coincide with the onset of acceleration. The left edge of the box is chosen to lie at \(x = 0\) in these coordinates. If the proper acceleration of the left edge is \(a\), the transformation between Minkowski and Rindler coordinates is given by
\begin{align}
    \label{TXtx}
    T &= \frac{1}{a} e^{a x} \sinh(at)\,, \\
    X &= \frac{1}{a} e^{a x} \cosh(at) - \frac{1}{a}\,. \label{TXtx2}
\end{align}

Since the box maintains a constant proper length \(L\), its right edge must also follow a trajectory of constant \(x\) in Rindler coordinates. To determine the Rindler coordinate \(x = l\) corresponding to the right edge, we compare the positions of both edges at the transition point \(T = t = 0\). At this instant, the Minkowski position of the right edge is \(X = L\), and using Eq.~\eqref{TXtx2}, we find
\begin{equation}
    \label{lL}
    l = \frac{1}{a} \ln(1 + aL)\,.
\end{equation}

This setup ensures that the box remains rigid in its instantaneous rest frame throughout its motion, allowing us to consistently define field modes and analyze their evolution across the inertial and accelerated regimes.

\subsection{Mode expansions}

We consider a massless, free scalar field \(\phi\) confined within a one-dimensional box of fixed proper length \(L\). The field satisfies Dirichlet boundary conditions, vanishing at both ends of the box. In the inertial regime, it is convenient to work in Minkowski coordinates \((T, X)\), where the box is at rest and the left edge is located at \(X = 0\). The mode expansion of the field in this frame is given by
\begin{equation}
    \label{phiTX}
    \phi_{\text{in}}(T, X) = \sum_{n=1}^{\infty} \frac{1}{\sqrt{n\pi}} \sin\left(n\pi\frac{X}{L}\right) \left( b_n e^{-i n\pi \frac{T}{L}} + b_n^\dagger e^{i n\pi \frac{T}{L}} \right)\,,
\end{equation}
where \(b_n\) and \(b_n^\dagger\) denote the annihilation and creation operators for the \(n\)-th mode, respectively. The normalization is chosen such that the canonical equal-time commutation relations,
\[
[\phi(T,X), \partial_T \phi(T,X')] = i \delta(X - X')
\]
imply the standard algebra for the mode operators,
\[
[b_n, b_m^\dagger] = \delta_{nm}\,, \quad [b_n, b_m] = [b_n^\dagger, b_m^\dagger] = 0\,.
\]
Physically, the field configuration corresponds to standing wave patterns trapped between perfectly reflecting boundaries. The discreteness of allowed wavenumbers arises from the confinement, and each normal mode behaves as an independent harmonic oscillator in the quantized theory.

Once the box begins to accelerate uniformly, it is natural to adopt Rindler coordinates \((t, x)\), in which the left edge of the box is located at \(x = 0\). The metric in these coordinates is conformally flat, and the equation of motion for the scalar field retains its form. The field modes defined with respect to the Rindler time \(t\) therefore describe oscillations as perceived by a comoving accelerated observer. Since the box maintains constant proper length, the right edge is fixed at \(x = l\), where \(l\) is given by Eq.~\eqref{lL}. The boundary conditions remain Dirichlet, and the mode expansion in the accelerated frame becomes
\begin{equation}
    \label{phitx}
    \phi_{\text{ac}}(t, x) = \sum_{n=1}^{\infty} \frac{1}{\sqrt{n\pi}} \sin\left(n\pi\frac{x}{l}\right) \left( c_n e^{-i n\pi \frac{t}{l}} + c_n^\dagger e^{i n\pi \frac{t}{l}} \right)\,,
\end{equation}
where \(c_n\) and \(c_n^\dagger\) are the annihilation and creation operators associated with the accelerated modes. As in the inertial case, these operators satisfy the canonical commutation relations
\begin{equation}
    [c_n, c_m^\dagger] = \delta_{nm}\,, \quad [c_n, c_m] = [c_n^\dagger, c_m^\dagger] = 0\,.
\end{equation}

The similarity in structure between Eqs.~\eqref{phiTX} and \eqref{phitx} reflects the conformal nature of the Rindler metric in \(1+1\) dimensions and the preservation of boundary conditions under the transition. Because the notions of positive frequency in the inertial and accelerated frames are defined with respect to different timelike Killing vectors, the associated annihilation and creation operators differ. Consequently, the field modes in the two frames are related through a Bogoliubov transformation, which encodes the redistribution of excitations due to acceleration.

\subsection{Continuity and Differentiability Conditions}

To consistently describe the evolution of the scalar field across the transition from inertial to accelerated motion at \(T = t = 0\), we require the field to be both continuous and differentiable at this instant. These matching conditions allow us to relate the mode expansions in the inertial and accelerated regimes, given by Eqs.~\eqref{phiTX} and \eqref{phitx}, respectively. In particular, they yield expressions for the accelerated mode operators \(c_n\) and \(c_n^\dagger\) in terms of the inertial operators \(b_n\) and \(b_n^\dagger\), thereby determining the Bogoliubov coefficients for the system.

Using the coordinate transformation in Eq.~\eqref{TXtx} and inverting Eq.~\eqref{lL}, we express the inertial field \(\phi_{\text{in}}\) in terms of Rindler coordinates \((t,x)\) as
\begin{equation}
    \phi_{\text{in}}(t,x) = \sum_{n=1}^{\infty} \frac{1}{\sqrt{n\pi}} \sin\left( n\pi \frac{e^{a x} \cosh(at) - 1}{e^{a l} - 1} \right) \left( b_n e^{-i n\pi \frac{e^{a x} \sinh(at)}{e^{a l} - 1}} + b_n^\dagger e^{i n\pi \frac{e^{a x} \sinh(at)}{e^{a l} - 1}} \right)\,.
\end{equation}

\subsubsection{Continuity Condition}

The continuity of the field at \(t = 0\) requires
\begin{equation}
    \phi_{\text{in}}(0,x) = \phi_{\text{ac}}(0,x)\,.
\end{equation}
Evaluating both sides at \(t = 0\), we obtain
\begin{equation}
    \sum_{m=1}^{\infty} \frac{1}{\sqrt{m\pi}} \sin\left( m\pi \frac{x}{l} \right) \left( c_m + c_m^\dagger \right) = \sum_{m=1}^{\infty} \frac{1}{\sqrt{m\pi}} \sin\left( m\pi \frac{e^{a x} - 1}{e^{a l} - 1} \right) \left( b_m + b_m^\dagger \right)\,.
\end{equation}
Multiplying both sides by \(\frac{1}{l} \sin\left( n\pi \frac{x}{l} \right)\) and integrating over \(x \in [0, l]\), we use the orthogonality relation
\begin{equation}
    \label{ortho}
    \int_0^l \frac{dx}{l} \sin\left( n\pi \frac{x}{l} \right) \sin\left( m\pi \frac{x}{l} \right) = \frac{1}{2} \delta_{nm}
\end{equation}
to isolate the \(n^\text{th}\) mode on the left-hand side. This yields
\begin{equation}
    \label{q}
    \frac{1}{2} \left( c_n + c_n^\dagger \right) = \sum_{m=1}^{\infty} I_{nm} \left( b_m + b_m^\dagger \right)\,,
\end{equation}
where the overlap integral \(I_{nm}\) is defined as
\begin{equation}
    I_{nm} = \sqrt{\frac{n}{m}} \int_0^l \frac{dx}{l} \sin\left( n\pi \frac{x}{l} \right) \sin\left( m\pi \frac{e^{a x} - 1}{e^{a l} - 1} \right)\,.
\end{equation}

Introducing the substitution \(z = \frac{e^{a x} - 1}{a}\), and recalling that \(l = \frac{1}{a} \ln(1 + aL)\), we can recast the integral as
\begin{equation}
    \label{I}
    I_{nm} = \sqrt{\frac{n}{m}} \frac{a}{\ln(1 + aL)} \int_0^L \frac{dz}{1 + a z} \sin\left( n\pi \frac{\ln(1 + a z)}{\ln(1 + aL)} \right) \sin\left( m\pi \frac{z}{L} \right)\,.
\end{equation}

\subsubsection{Differentiability Condition}

Next, we impose differentiability of the field at \(t = 0\), requiring
\begin{equation}
    \left. \partial_t \phi_{\text{in}}(t,x) \right|_{t=0} = \left. \partial_t \phi_{\text{ac}}(t,x) \right|_{t=0}\,.
\end{equation}
Evaluating both sides, we obtain
\begin{equation}
    \sum_{m=1}^{\infty} \sqrt{\frac{m}{\pi}} \frac{1}{l} \sin\left( m\pi \frac{x}{l} \right) \left( c_m - c_m^\dagger \right) = \sum_{m=1}^{\infty} \sqrt{\frac{m}{\pi}} \frac{a e^{a x}}{e^{a l} - 1} \sin\left( m\pi \frac{e^{a x} - 1}{e^{a l} - 1} \right) \left( b_m - b_m^\dagger \right)\,.
\end{equation}
Multiplying both sides by \(\frac{1}{l} \sin\left( n\pi \frac{x}{l} \right)\) and integrating over \(x \in [0, l]\), as before using the orthogonality relation of Eq.(\ref{ortho}) to isolate the \(n^\text{th}\) mode on the left-hand side, we obtain
\begin{equation}
    \label{p}
    \frac{1}{2} \left( c_n - c_n^\dagger \right) = \sum_{m=1}^{\infty} J_{nm} \left( b_m - b_m^\dagger \right)\,,
\end{equation}
where the coefficient \(J_{nm}\) is given by
\begin{equation}
    J_{nm} = \sqrt{\frac{m}{n}} \frac{a l}{e^{a l} - 1} \int_0^l \frac{dx}{l} e^{a x} \sin\left( n\pi \frac{x}{l} \right) \sin\left( m\pi \frac{e^{a x} - 1}{e^{a l} - 1} \right)\,.
\end{equation}

Using the same substitution \(z = \frac{e^{a x} - 1}{a}\), we can rewrite the integral as
\begin{equation}
    \label{J}
    J_{nm} = \sqrt{\frac{m}{n}} \frac{1}{L} \int_0^L dz \sin\left( n\pi \frac{\ln(1 + a z)}{\ln(1 + aL)} \right) \sin\left( m\pi \frac{z}{L} \right)\,.
\end{equation}

Eqs.(\ref{q}) and (\ref{p}) together define the Bogoliubov transformation relating the inertial and accelerated mode operators. Both the coefficients \(I_{nm}\) and \(J_{nm}\) are dimensionless and depend only on the dimensionless parameter $aL$, which characterizes the strength of acceleration relative to the box size. They encode how the inertial field modes decompose in the accelerated basis, with non-vanishing off-diagonal elements signifying mode coupling induced by the transition to accelerated motion. This coupling reflects the fact that the notion of vacuum for an accelerated observer differs from that of an inertial one, thereby providing the framework for analyzing field excitations and particle creation in the subsequent sections.

\section{The Bogoliubov Coefficients}

We now turn to the evaluation of the Bogoliubov transformation connecting the inertial and accelerated field modes within the rigid box.

Having established the continuity and differentiability conditions across the transition at \(T = t = 0\), we can express the mode operators in the accelerated frame as linear combinations of those in the inertial frame. Specifically, the annihilation and creation operators \(c_n\) and \(c_n^\dagger\) are related to \(b_n\) and \(b_n^\dagger\) via the Bogoliubov transformation:
\begin{align}
    \label{cb1}
    c_n &= \sum_{m=1}^\infty \left( \alpha_{nm} b_m - \beta_{nm} b_m^\dagger \right)\,, \\
    \label{cb2}
    c_n^\dagger &= \sum_{m=1}^\infty \left( \beta^*_{nm} b_m - \alpha^*_{nm} b_m^\dagger \right)\,.
\end{align}

From the continuity and differentiability conditions given in Eq.(\ref{q}) and Eq.(\ref{p}), we can directly read off the Bogoliubov coefficients \(\alpha_{nm}\) and \(\beta_{nm}\) as
\begin{align}
    \label{ab}
    \begin{rcases}
        \alpha_{nm} \\
        \beta_{nm}
    \end{rcases}
    = \pm I_{nm} + J_{nm}\,,
\end{align}
where \(I_{nm}\) and \(J_{nm}\) are the overlap integrals defined in Eq.(\ref{I}) and Eq.(\ref{J}), respectively.

Before combining \(I_{nm}\) and \(J_{nm}\), it is useful to express them in alternative forms. Consider the integrand in Eq.(\ref{I}). A part can be written as a total derivative.
\begin{equation}
    n\pi \frac{a}{\ln(1 + aL)} \frac{1}{1 + a z} \sin\left( n\pi \frac{\ln(1 + a z)}{\ln(1 + aL)} \right) = -\frac{d}{dz} \left[ \cos\left( n\pi \frac{\ln(1 + a z)}{\ln(1 + aL)} \right) \right]
\end{equation}
This allows us to perform integration by parts on Eq.(\ref{I}). The boundary terms vanish, yielding
\begin{equation}
    \label{I1}
    I_{nm} = \sqrt{\frac{m}{n}} \frac{1}{L} \int_0^L dz \cos\left( n\pi \frac{\ln(1 + a z)}{\ln(1 + aL)} \right) \cos\left( m\pi \frac{z}{L} \right)\,.
\end{equation}
This manipulation highlights that the acceleration enters the overlap integrals through the logarithmic deformation of the spatial argument, which modifies the phase structure of each mode. Substituting Eq.(\ref{I1}) and Eq.(\ref{J}) into Eq.(\ref{ab}), we obtain a compact expression for the Bogoliubov coefficients
\begin{align}
    \label{I_m/n}
    \begin{rcases}
        \alpha_{nm} \\
        \beta_{nm}
    \end{rcases}
    = \pm \sqrt{\frac{m}{n}} \frac{1}{L} \int_0^L dz \cos\left( n\pi \frac{\ln(1 + a z)}{\ln(1 + aL)} \mp m\pi \frac{z}{L} \right)
\end{align}

Alternatively, we may integrate Eq.(\ref{J}) by parts. Using
\begin{equation}
    m\pi \frac{1}{L} \sin\left( m\pi \frac{z}{L} \right) = -\frac{d}{dz} \left[ \cos\left( m\pi \frac{z}{L} \right) \right]\,,
\end{equation}
and noting that the boundary terms vanish, we obtain
\begin{equation}
    \label{J1}
    J_{nm} = \sqrt{\frac{n}{m}} \frac{a}{\ln(1 + aL)} \int_0^L dz \cos\left( n\pi \frac{\ln(1 + a z)}{\ln(1 + aL)} \right) \cos\left( m\pi \frac{z}{L} \right)\,.
\end{equation}

Combining Eq.(\ref{J1}) with Eq.(\ref{I}) in Eq.(\ref{ab}) yields an alternative expression for the Bogoliubov coefficients:
\begin{align}
    \label{I_n/m}
    \begin{rcases}
        \alpha_{nm} \\
        \beta_{nm}
    \end{rcases}
    = \sqrt{\frac{n}{m}} \frac{a}{\ln(1 + aL)} \int_0^L \frac{dz}{1 + a z} \cos\left( n\pi \frac{\ln(1 + a z)}{\ln(1 + aL)} \mp m\pi \frac{z}{L} \right)\,.
\end{align}
The two alternative representations, (\ref{I_m/n}) and (\ref{I_n/m}), emphasize different asymptotic regimes: the former is more convenient for large \(n\), while the latter simplifies for large \(m\), corresponding to high-frequency inertial or accelerated modes respectively. That they agree has been numerically checked as well.

\subsection{Proper Time Dependence}

Until now, the coefficients have been evaluated taking instant the acceleration begins as the zero of time. To understand a more general transition occurring at some specified arbitrary proper time, we next incorporate the phase dependence explicitly.

It is worth noting that the Bogoliubov coefficients derived above are real. This is a consequence of our choice to set \(T = 0\) and \(t = 0\) as the instant the box begins accelerating. To generalize, let us introduce a proper time parameter \(\tau\) associated with the left edge of the box, such that the transition occurs at \(\tau = \tau_0\). In the inertial regime (\(\tau \leq \tau_0\)), we have \(T = \tau - \tau_0\), while in the accelerated regime (\(\tau > \tau_0\)), we have \(t = \tau - \tau_0\). The temporal dependence of the \(n^\text{th}\) mode in the inertial frame is given by
\[
b_n e^{-i \frac{n\pi}{L} T} = b_n e^{-i \frac{n\pi}{L} (\tau - \tau_0)}\,,
\]
and similarly, in the accelerated frame,
\[
c_n e^{-i \frac{n\pi}{l} t} = c_n e^{-i \frac{n\pi}{l} (\tau - \tau_0)}\,.
\]
Thus, the appropriate annihilation and creation operators in each frame are modified by phase factors:
\begin{align*}
    b_n &\rightarrow b_n e^{i n\pi \frac{\tau_0}{L}}\,, & b_n^\dagger &\rightarrow b_n^\dagger e^{-i n\pi \frac{\tau_0}{L}}\,, \\
    c_n &\rightarrow c_n e^{i n\pi \frac{\tau_0}{l}}\,, & c_n^\dagger &\rightarrow c_n^\dagger e^{-i n\pi \frac{\tau_0}{l}}\,.
\end{align*}

Accordingly, the Bogoliubov transformation appropriate for this choice of proper time becomes
\begin{align}
    \alpha^{(\tau_0)}_{nm} &= e^{i n\pi \frac{\tau_0}{l}} \alpha_{nm} e^{-i m\pi \frac{\tau_0}{L}}\,, \\
    \label{timefactor}
    \beta^{(\tau_0)}_{nm} &= e^{i n\pi \frac{\tau_0}{l}} \alpha_{nm} e^{i m\pi \frac{\tau_0}{L}}\,.
\end{align}

These expressions allow us to track the phase evolution of the modes relative to a general choice of transition time \(\tau_0\), and will be useful when considering time-dependent acceleration profiles.

\subsection{Exact Solution in Terms of the Incomplete Gamma Functions}

While the integral expressions of Eq.(\ref{I_m/n}) and Eq.(\ref{I_n/m}) obtained for the Bogoliubov coefficients are compact, they obscure the analytic structure of the coefficients. To make this structure explicit and facilitate asymptotic analysis, it is useful to obtain a closed-form representation.

To proceed, we express the cosine function as the real part of a complex exponential:
\begin{align}
    \begin{rcases}
        \alpha_{nm} \\
        \beta_{nm}
    \end{rcases}
    &= \Re\left[ \sqrt{\frac{n}{m}} \frac{a}{\ln(1 + aL)} \int_0^L \frac{dz}{1 + a z} e^{-i n\pi \frac{\ln(1 + a z)}{\ln(1 + aL)}} e^{\pm i m\pi \frac{z}{L}} \right] \\
    &= \Re\left[ \sqrt{\frac{n}{m}} \frac{a}{\ln(1 + aL)} \int_0^L dz\, (1 + a z)^{\frac{-i n\pi}{\ln(1 + aL)} - 1} e^{\pm i m\pi \frac{z}{L}} \right] \\
    &= \Re\left[ \sqrt{\frac{n}{m}} \frac{1}{\ln(1 + aL)} \int_1^{1 + aL} dw\, w^{\frac{-i n\pi}{\ln(1 + aL)} - 1} e^{\pm i m\pi \frac{w - 1}{aL}} \right] \\
    &= \Re\left[ \sqrt{\frac{n}{m}} \frac{1}{\ln(1 + aL)} \int_{\mp i m\pi / aL}^{\mp i m\pi / aL \mp i m\pi} \frac{\pm i aL}{m\pi} dq\, \left( \frac{\pm i aL}{m\pi} q \right)^{\frac{-i n\pi}{\ln(1 + aL)} - 1} e^{-q} e^{\mp i \frac{m\pi}{aL}} \right] \\
    &= \Re\left[ \sqrt{\frac{n}{m}} \frac{1}{\ln(1 + aL)} \left( \mp i \frac{m\pi}{aL} \right)^{\frac{i n\pi}{\ln(1 + aL)}} e^{\mp i \frac{m\pi}{aL}} \int_{\mp i m\pi / aL}^{\mp i m\pi / aL \mp i m\pi} dq\, q^{-i \frac{n\pi}{\ln(1 + aL)} - 1} e^{-q} \right]
\end{align}

Here, we have substituted \(w = 1+az\) and \(q = \frac{m\pi}{aL}w\). In the final form, we identify the integral as the incomplete gamma function, defined by
\begin{equation}
    \label{gammainc}
    \gamma(s, u, v) = \int_u^v dq\, q^{s - 1} e^{-q}\,.
\end{equation}

Using the identity \((\mp i)^i = e^{\pm \pi / 2}\), we obtain a closed-form expression for the Bogoliubov coefficients:
\begin{align}
 \label{ab_exact2}
    \begin{rcases}
        \alpha_{nm} \\
        \beta_{nm}
    \end{rcases}
    &= \Re\Bigg[ \sqrt{\frac{n}{m}} \frac{1}{\ln(1 + aL)} e^{\pm \frac{n\pi^2}{2 \ln(1 + aL)}} e^{i \frac{n\pi}{\ln(1 + aL)} \ln\left( \frac{m\pi}{aL} \right)} e^{\mp i \frac{m\pi}{aL}} \nonumber \\
    &\quad\quad\quad\quad\quad\quad\quad\quad \times \gamma\left( -i \frac{n\pi}{\ln(1 + aL)}, \mp i \frac{m\pi}{aL}, \mp i \frac{m\pi}{aL} \mp i m\pi \right) \Bigg]
\end{align}
The agreement of this equation with Eq.(\ref{I_n/m}) has been numerically checked as well.

It is noteworthy to recast this expression in terms of the mode frequencies. The frequency of the \(n^\text{th}\) mode in the accelerated box is
\begin{equation}
    \label{freqleft}
    \Omega_n \coloneqq \frac{n\pi}{l} = \frac{n\pi a}{\ln(1 + aL)}\,,
\end{equation}
while the frequency of the \(m^\text{th}\) mode in the inertial box is $\omega_m \coloneqq m\pi/L$.
In terms of these variables, Eq.(\ref{ab_exact2}) becomes
\begin{align}
    \label{ab_exact}
    \begin{rcases}
        \alpha_{nm} \\
        \beta_{nm}
    \end{rcases}
    &= \frac{1}{\sqrt{aL \ln(1 + aL)}} \sqrt{\frac{\Omega_n}{\omega_m}} e^{\pm \frac{\pi}{2} \frac{\Omega_n}{a}} \Re\Bigg[ e^{i \frac{\Omega_n}{a} \ln\left( \frac{\omega_m}{a} \right)} e^{\mp i \frac{\omega_m}{a}} \gamma\left( -i \frac{\Omega_n}{a}, \mp i \frac{\omega_m}{a}, \mp i \frac{\omega_m}{a}(1 + aL) \right) \Bigg]
\end{align}
This expression makes transparent the dependence of the Bogoliubov coefficients on the ratio of mode frequencies to acceleration, showing how the confinement modifies the characteristic Unruh-type frequency scaling. Further, this form bears a striking resemblance to the Bogoliubov transformation for a scalar field propagating freely in Minkowski spacetime, when comparing inertial and uniformly accelerated observers. In the latter case, the Bogoliubov coefficients are given by~\cite{Mukhanov_Winitzki_2007}
\begin{align}
    \begin{rcases}
        \alpha_{nm} \\
        \beta_{nm}
    \end{rcases}
    = \pm \frac{1}{2\pi a} \sqrt{\frac{\Omega_n}{\omega_m}} e^{\pm \frac{\pi}{2} \frac{\Omega_n}{a}} e^{i \frac{\Omega_n}{a} \ln\left( \frac{\omega_m}{a} \right)} \Gamma\left( -i \frac{\Omega_n}{a} \right)\,.
\end{align}
These coefficients are non-invertible and lead to a thermal state of the field in the accelerated frame which is a manifestation of the Unruh effect. Comparing these with the coefficients obtained in Eq.(\ref{ab_exact2}) leads to few notable key differences. The Bogoliubov transformation derived here is real, invertible, and involves the incomplete gamma function \(\gamma(s,u,v)\) rather than the complete gamma function \(\Gamma(s)\). This distinction reflects the finite spatial extent of the system and the non-stationary nature of the transition, which together prevent the emergence of a thermal particle distribution. Further implications for the nature and spectrum of particle creation shall be explored in the following sections.

\subsection{The Excitation Spectrum}

Having determined the Bogoliubov coefficients relating the inertial and accelerated modes, we next look at the physical consequences of acceleration on the quantum field. In particular, we analyze the spectrum of excitations induced when the box begins to accelerate, signifying particle creation in a confined system. Even though the field remains spatially confined, the sudden transition to uniform acceleration effectively acts as a time-dependent perturbation, redistributing vacuum fluctuations into real excitations.

We assume the field is initially in its vacuum state \(\ket{S}\) during the inertial phase defined by
\begin{align}
    \label{vac}
    b_n \ket{S} = 0 \qquad \forall n\,,
\end{align}
where \(b_n\) are the annihilation operators as defined in Eq.(\ref{phiTX}) associated with the inertial modes.

Once the box accelerates, the relevant mode operators are \(c_n\) and \(c_n^\dagger\), corresponding to the accelerated frame. The expected number of excitations in the \(n^\text{th}\) accelerated mode, when the field is in the state \(\ket{S}\), is given by
\begin{equation}
    P_n = \bra{S} c_n^\dagger c_n \ket{S}\,.
\end{equation}

Using the Bogoliubov transformation from Eq.(\ref{cb1}) and Eq.(\ref{cb2}), along with the vacuum condition in Eq.(\ref{vac}), we find that
\begin{equation}
    \label{P}
    P_n = \sum_{m=1}^{\infty} \beta_{nm} \beta_{nm}^*\,.
\end{equation}
This quantity \(P_n\) represents the excitation spectrum of the field in the accelerated box, assuming the field was initially unexcited in the inertial vacuum. Physically, it quantifies the number of particles created in each mode due to the sudden acceleration of the box.

\begin{figure}[h]
    \centering
    \includegraphics[width=0.7\linewidth]{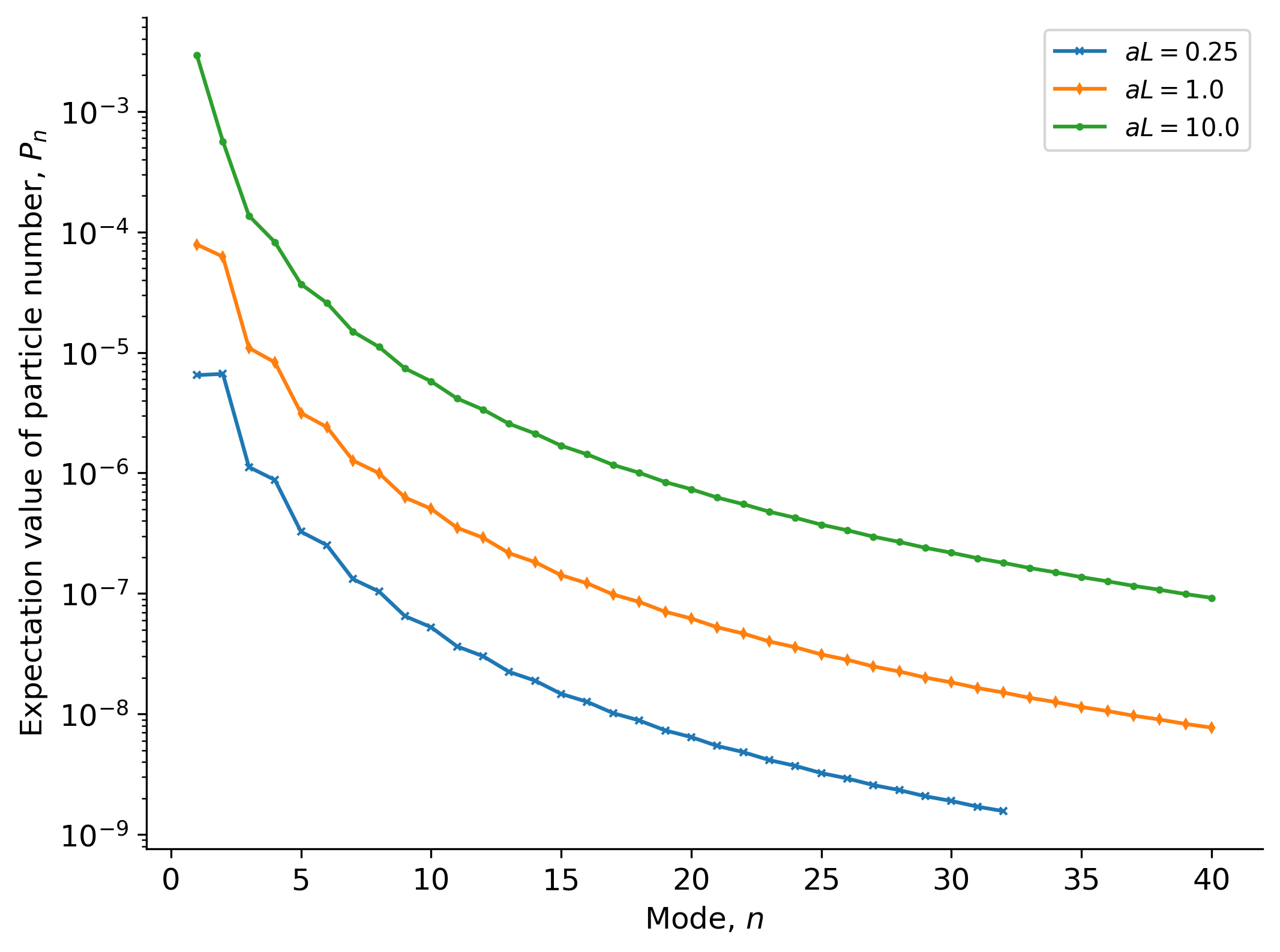}
    \caption{The spectra of particles created as the box starts accelerating, plotted for different values of \(aL\).}
    \label{fig:spectra}
\end{figure}

Figure~\ref{fig:spectra} shows the excitation spectrum \(P_n\) computed for various values of the dimensionless parameter \(aL\). The plot is generated by numerically evaluating Eq.(\ref{P}) using the closed-form expression for \(\beta_{nm}\) given in Eq.(\ref{ab_exact}). As expected, larger values of \(aL\) correspond to greater particle production, reflecting the stronger influence of acceleration.

For each fixed value of \(aL\), the excitation spectrum exhibits a smooth decay with increasing mode number \(n\), indicating that higher-frequency modes are less populated. Interestingly, for small values of \(aL\), particularly in the low-mode regime, the spectrum displays an alternating pattern that favors even modes over adjacent odd ones. This modulation is superposed on the overall decay and suggests a subtle structure in the mode mixing induced by acceleration.

\section{Small Acceleration Approximation}

Before investigating the general case, we begin by analyzing the Bogoliubov transformation in the limit of small acceleration, retaining terms only up to first order in \(a\). This regime corresponds to situations where the box is gently accelerated, so the field modes are only slightly perturbed from their inertial form. Studying this limit gives insight into the onset of particle creation and the role of mode mixing.

\subsection{Bogoliubov Coefficients}

To calculate the Bogoliubov Coefficients in the limit of small acceleration, it is convenient to work with the expression in Eq.(\ref{I_m/n}). Expanding the logarithmic terms to first order, we find
\begin{align}
    \frac{\ln(1 + a z)}{\ln(1 + a L)} &\approx \frac{a z - \frac{1}{2} a^2 z^2}{a L - \frac{1}{2} a^2 L^2} \approx \frac{z}{L} \left(1 + \frac{1}{2} a (L - z)\right)\,.
\end{align}

Substituting this into Eq.(\ref{I_m/n}), the Bogoliubov coefficients become
\begin{align}
    \begin{rcases}
        \alpha_{nm} \\
        \beta_{nm}
    \end{rcases}
    &\approx \pm \sqrt{\frac{m}{n}} \frac{1}{L} \int_0^L dz \cos\left( n\pi \frac{z}{L} \left(1 + \frac{1}{2} a (L - z)\right) \mp m\pi \frac{z}{L} \right) \\
    &\approx \pm \sqrt{\frac{m}{n}} \frac{1}{L} \int_0^L dz \left[ \cos\left( (n\pi \mp m\pi) \frac{z}{L} \right) - n\pi a \frac{z(L - z)}{L} \sin\left( (n\pi \mp m\pi) \frac{z}{L} \right) \right]\,.
\end{align}

The zeroth-order term corresponds to the case where the box remains inertial, and the mode functions are unchanged.
\begin{align}
    \begin{rcases}
        \alpha^{(0)}_{nm} \\
        \beta^{(0)}_{nm}
    \end{rcases}
    = \pm \sqrt{\frac{m}{n}} \int_0^1 dy \cos\left( (n\pi \mp m\pi) y \right) = 
    \begin{cases}
        \delta_{nm} \\
        0
    \end{cases}
\end{align}
Here we have introduced the substitution \(y = z / L\) for convenience. This is a consistency check that, without acceleration, the observer would see exactly the same vacuum as the inertial one, with no excitations. The first-order contribution is given by
\begin{align}
    \begin{rcases}
        \alpha^{(1)}_{nm} \\
        \beta^{(1)}_{nm}
    \end{rcases}
    = \mp \sqrt{n m} \pi a L \int_0^1 dy\, y(1 - y) \sin\left( (n\pi \mp m\pi) y \right)\,.
\end{align}
The presence of the factor $y(1-y)$ indicates  that the contribution from acceleration is mostly near the center of the box and vanishes at the edges, perhaps reflecting the rigidity of the box boundaries. Moreover, the sine term indicates that particle creation arises primarily from mode interference and only certain pairs of modes mix, producing the alternating pattern observed in the low $aL$ spectrum. Evaluating the integral using integration by parts, we obtain
\begin{align}
    \label{alphasmalla}
    \alpha^{(1)}_{nm} &= 
    \begin{cases}
        -aL \frac{2}{\pi^2} \frac{\sqrt{n m}}{(n - m)^3} & \text{if } n + m \text{ is odd} \\
        0 & \text{if } n + m \text{ is even}
    \end{cases} \\
    \label{betasmalla}
    \beta^{(1)}_{nm} &= 
    \begin{cases}
        aL \frac{2}{\pi^2} \frac{\sqrt{n m}}{(n + m)^3} & \text{if } n + m \text{ is odd} \\
        0 & \text{if } n + m \text{ is even}
    \end{cases}
\end{align}
These expressions have been found previously \cite{Bruschi2012Voyage} and are in agreement with the literature.

\subsection{Excitation Spectrum}

Using the approximate expression Eq.(\ref{betasmalla}) for small acceleration in Eq.(\ref{P}), the expectation value of the number of particles produced in the \(n^\text{th}\) mode, assuming the field was initially in the inertial vacuum, is given by
\begin{align}
    \label{lowaL}
    P_n = a^2 L^2 \frac{4}{\pi^4} \sum_{\substack{m = 1 \\ n + m = \text{odd}}}^{\infty} \frac{n m}{(n + m)^6} =\frac{4a^2L^2}{\pi^2}\begin{cases}
        \sum_{m=1}^\infty \frac{n(2m-1)}{n+2m-1} & \text{if } n \text{ is even}\\[6 pt]\sum_{m=1}^\infty \frac{n(2m)}{n+2m} & \text{if } n \text{ is odd}
    \end{cases}\,.
\end{align}
The structure of the summation differs significantly between even and odd values of \(n\), leading to distinct excitation patterns, namely
\begin{align}
    \label{expsum}
    P_n = \frac{4 a^2 L^2}{\pi^4} \times
    \begin{cases}
        \frac{n}{(n + 1)^3} + \frac{3n}{(n + 3)^3} + \frac{5n}{(n + 5)^3} + \dots & \text{if } n \text{ is even} \\
        \frac{2n}{(n + 2)^3} + \frac{4n}{(n + 4)^3} + \frac{6n}{(n + 6)^3} + \dots & \text{if } n \text{ is odd}
    \end{cases}
\end{align}
The quadratic dependence on $aL$ reveals that the leading order particle production rate grows with the square of the acceleration,  similar to other non-inertial effects such as the Larmor radiation~\cite{Larmor1897}. However, the discrete boundary conditions modify the spectrum by imposing mode-dependent selection rules.

\begin{figure}[h]
    \centering
    \includegraphics[width=0.7\linewidth]{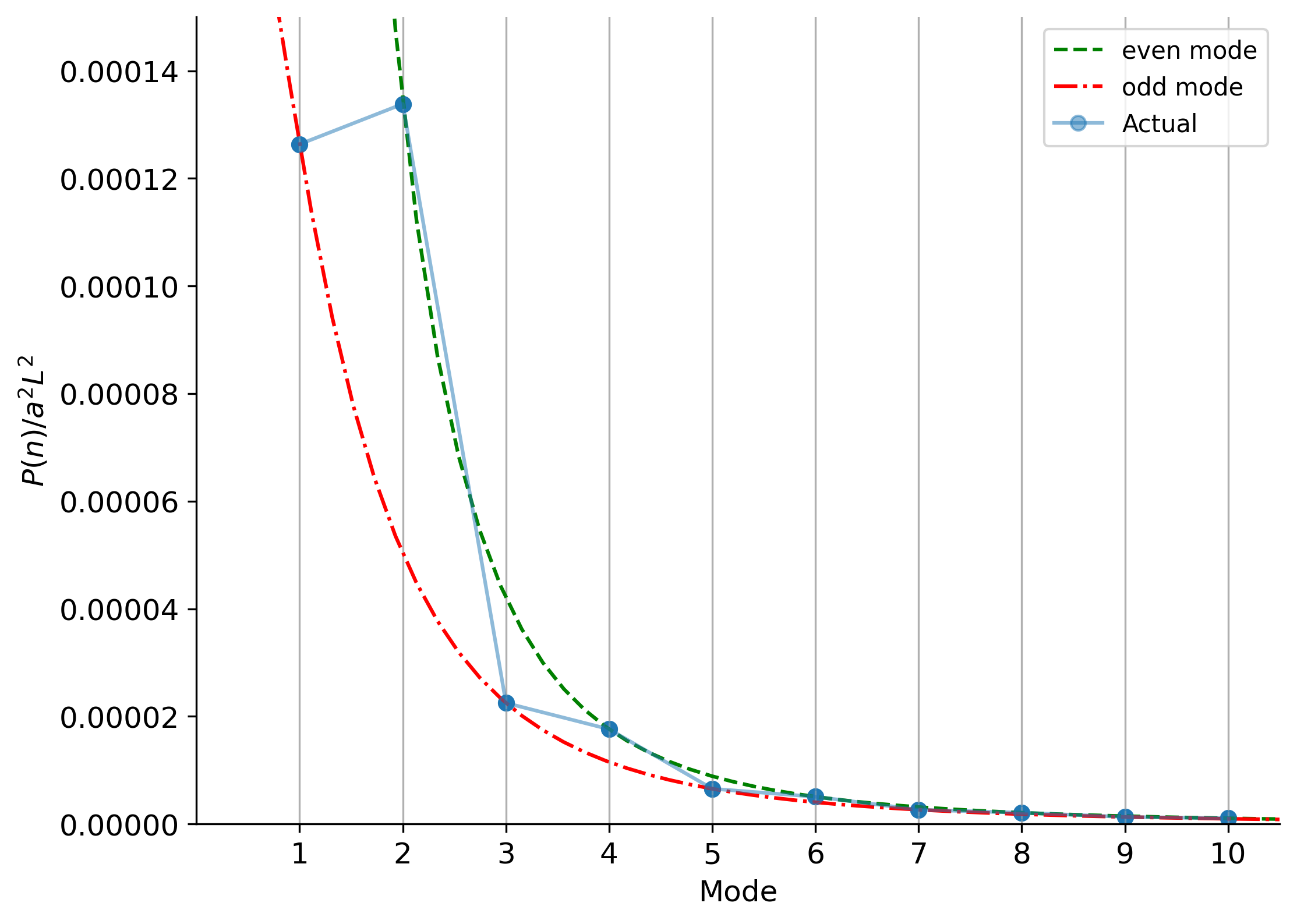}
    \caption{A plot showing the excitation spectrum in the low \(aL\) limit.}
    \label{fig:lowaL}
\end{figure}

Figure~\ref{fig:lowaL} shows the excitation spectrum \(P_n\) in the low-\(aL\) regime, along with separate curves for even and odd modes. The alternating pattern favoring even modes, observed in Fig.~\ref{fig:spectra}, is clearly explained by the structure of Eq.(\ref{expsum}).

\subsection{Large-\(n\) Decay}

To understand the asymptotic behavior of the excitation spectrum for large mode numbers, we analyze Eq.(\ref{lowaL}) in the limit \(n \gg 1\). Introducing the variable \(\zeta = m / n\), the summation becomes a Riemann sum with step size \(2 / n\), which can be approximated by the integral
\begin{equation}
    \label{smallint}
    P_n \approx \frac{2 a^2 L^2}{\pi^4 n^3} \int_0^\infty d\zeta\, \frac{\zeta}{(1 + \zeta)^6}\,.
\end{equation}
Evaluating the integral yields the asymptotic form
\begin{equation}
    \label{largen_lowaL}
    P_n = \frac{1}{10 \pi^4} \frac{a^2 L^2}{n^3}\,.
\end{equation}
The result in Eq.(\ref{largen_lowaL}) is particularly significant. It applies to modes with wavelengths much smaller than the box length, i.e. the regime \(n \gg 1\). This is the standard limit in which discrete spectra are often approximated by continuous ones. The $1/n^3$ decay law indicates that high-frequency (short-wavelength) modes are relatively less sensitive to the effects of small acceleration. Physically, one may try to understand this as these modes oscillate too rapidly to respond to the comparatively slow change in the boundary’s trajectory, leading to a suppression of particle creation at large $n$. This feature ensures the total excitation energy remains finite and dominated by low frequency modes.

In the following section, we extend this analysis to the general case, beyond the first-order approximation in \(aL\).

\section{Excitation spectrum for arbitrary acceleration $a$ in the large \(n\) limit}

\begin{figure}[h]
    \centering
    \includegraphics[width=0.4\linewidth]{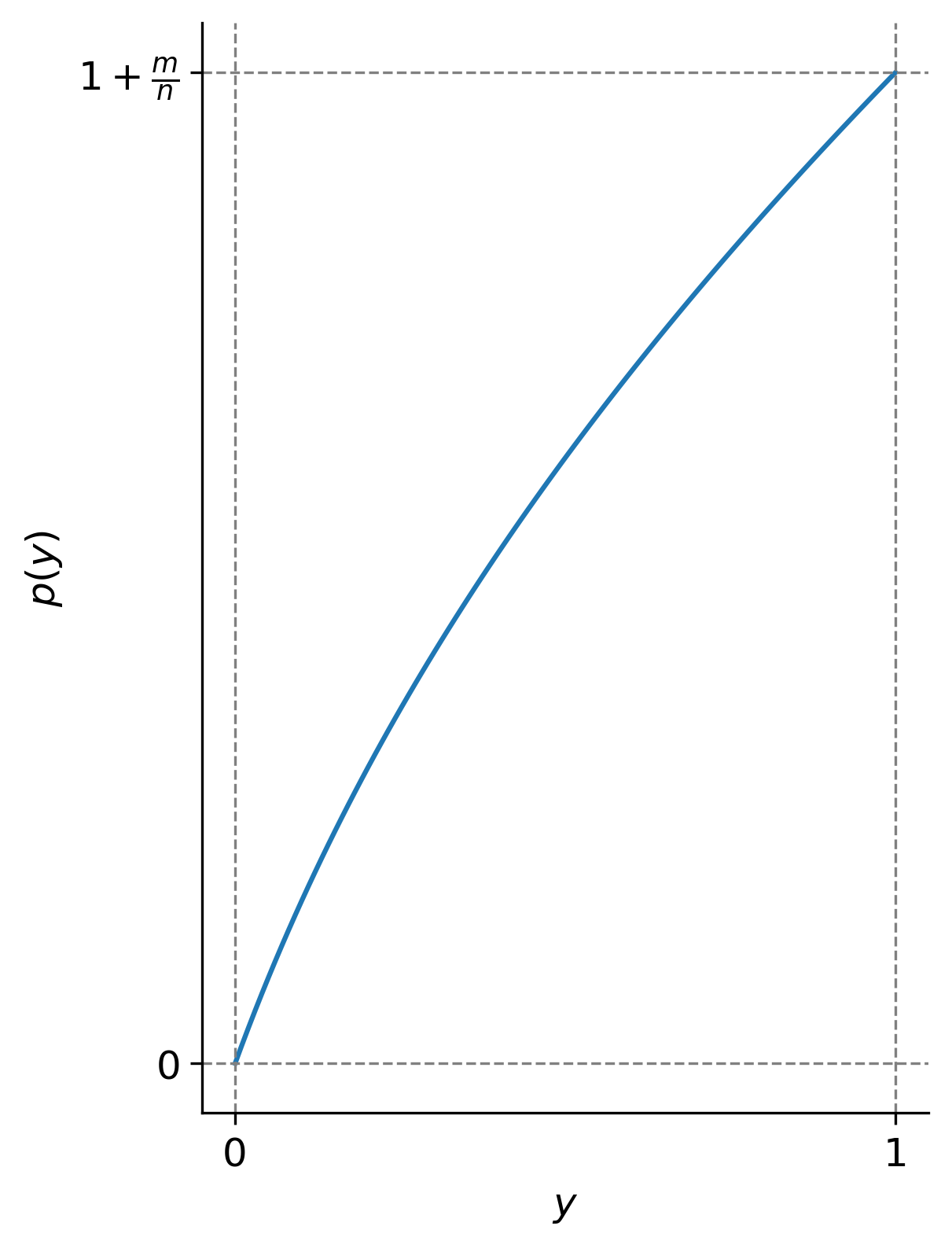}
    \caption{A generic plot of the function \(p(y)\).}
    \label{fig:py}
\end{figure}

We next analyze the behavior of excitations produced, for the general case of arbitrary acceleration, in modes with wavelengths much smaller than the size of the box, i.e., in the limit of large mode number \(n\). Our goal is to evaluate the Bogoliubov coefficient \(\beta_{nm}\), given exactly in Eq.(\ref{ab_exact}), in the asymptotic regime \(n \to \infty\). For this purpose, it is convenient to begin with the integral representation in Eq.(\ref{I_m/n}), which we recast as
\begin{align}
    \label{bln}
    \beta_{nm} &= -\sqrt{\frac{m}{n}} \int_0^1 dy\, \cos\left( n\pi p(y) \right)\,, \\
    p(y) &\coloneqq \frac{\ln(1 + aL y)}{\ln(1 + aL)} + \frac{m}{n} y\,.
\end{align}
We evaluate Eq.(\ref{bln}) in the limit \(n \to \infty\), holding the function \(p(y)\) fixed. The function \(p(y)\) is strictly monotonically increasing from \(0\) to \(1 + m/n\) as \(y\) varies from \(0\) to \(1\), as illustrated in Fig.~\ref{fig:py}. The phase \(n\pi p(y)\) thus undergoes \(n + m\) half-cycles over the interval, and we must treat the cases of even and odd \(n + m\) separately. The function $p(y)$ effectively encodes the nonlinear mapping between inertial and accelerated spatial coordinates.

Effectively, we are interested in evaluating the following oscillatory integral
\begin{align}
    \xi = \int_0^1 dy\, \cos\left( n\pi p(y) \right)\,,
\end{align}
where the integrand undergoes either an integer or half-integer number of oscillations depending on the parity of \(n + m\).

\subsection{Even \(n + m\)}

In this case, the integrand completes \((n + m)/2\) full oscillations over the interval \(y \in [0, 1]\). As \(n \to \infty\), the oscillations become increasingly dense, and each cycle spans a narrow region in \(y\). The \(r^\text{th}\) oscillation occurs as \(p\) ranges from \((2r - 2)/n\) to \(2r/n\). Using the inverse function \(y(p)\), the contribution from the \(r^\text{th}\) oscillation is
\begin{align}
    \delta \xi_r &= \int_{y\left( \frac{2r - 2}{n} \right)}^{y\left( \frac{2r}{n} \right)} dy\, \cos\left( n\pi p(y) \right) \\
    &= \int_{\frac{2r - 2}{n}}^{\frac{2r}{n}} dp\, \frac{dy}{dp} \cos(n\pi p)\,.
\end{align}
\(y(p)\), or its derivative, cannot be written explicitly in terms of its argument. However, as \(n \to \infty\), the above integration interval becomes infinitesimal, allowing us to Taylor expand \(\frac{dy}{dp}\) around the midpoint \(p_0 = (2r - 1)/n\) leading to
\begin{align}
    \delta \xi_r = \int_{\frac{2r - 2}{n}}^{\frac{2r}{n}} dp \left[ \left( \frac{dy}{dp} \right)_{p_0} + \left( \frac{d^2 y}{dp^2} \right)_{p_0} (p - p_0) + \frac{1}{2} \left( \frac{d^3 y}{dp^3} \right)_{p_0} (p - p_0)^2 + \dots \right] \cos(n\pi p)\,.
\end{align}
The first two terms vanish upon integration due to the symmetry of the cosine function. The leading non-zero contribution arises from the third term;
\begin{align}
    \delta \xi_r &\approx \left( \frac{d^3 y}{dp^3} \right)_{p_0} \frac{1}{2} \int_{\frac{2r - 2}{n}}^{\frac{2r}{n}} dp\, (p - p_0)^2 \cos(n\pi p) \\
    &= -\left( \frac{d^3 y}{dp^3} \right)_{p_0} \frac{1}{2 n^3 \pi^3} \int_{-\pi}^{\pi} dq\, q^2 \cos(q) \\
    &= \left( \frac{d^3 y}{dp^3} \right)_{p_0} \frac{2}{n^3 \pi^2}\,,
\end{align}
where we used the substitution \(q = n\pi(p - p_0)\). Summing over all oscillations, we obtain
\begin{align}
    \xi = \sum_{r = 1}^{\frac{n + m}{2}} \delta \xi_r = \frac{2}{n^3 \pi^2} \sum_{r = 1}^{\frac{n + m}{2}} \left( \frac{d^3 y}{dp^3} \right)_{\frac{2r - 1}{n}}\,.
\end{align}
For large \(n\), this sum can be approximated by an integral. Defining \(\chi = 2r/n\), we write
\begin{align}
    \xi &= \frac{2}{n^3 \pi^2} \sum_{\substack{\chi = 2/n \\ \text{step} = 2/n}}^{1 + \frac{m}{n}} \left( \frac{d^3 y}{dp^3} \right)_{\chi - \frac{1}{n}} \\
    &\approx \frac{1}{n^2 \pi^2} \int_0^{1 + \frac{m}{n}} d\chi\, \frac{d^3 y}{dp^3}(\chi) \\
    &= \frac{1}{n^2 \pi^2} \left( \frac{d^2 y}{dp^2}\left(1 + \frac{m}{n}\right) - \frac{d^2 y}{dp^2}(0) \right)\,.
\end{align}
Next we rewrite the unknown function $y(p)$ back in terms of its inverse $p(y)$ using
\begin{align}
    \label{d2ydp2}
    \frac{d^2 y}{dp^2} = \frac{d}{dp} \left( \frac{dy}{dp} \right) = \frac{1}{p'(y)} \frac{d}{dy} \left( \frac{1}{p'(y)} \right) = -\frac{p''(y)}{p'(y)^3}\,,
\end{align}
and noting that \(p(0) = 0\) and \(p(1) = 1 + m/n\), we obtain the leading order final result for even \(n + m\):
\begin{align}
    \label{evenxi}
    \xi = \frac{1}{n^2 \pi^2} \left( \frac{p''(0)}{p'(0)^3} - \frac{p''(1)}{p'(1)^3} \right) + \mathcal{O}\left( \frac{1}{n^3} \right)\,.
\end{align}
This procedure effectively replaces the rapidly oscillating integrand with a smoothed average over each cycle, allowing us to extract the leading-order behavior. The above expression shows that the leading contribution to the oscillatory integral depends entirely on the behavior of \(p(y)\) the end points of the interval.

\subsection{Odd \(n + m\)}

In the odd  \(n + m\) case, the integrand \(\cos\left(n\pi p(y)\right)\) completes \((n + m - 1)/2\) full oscillations over the interval \(y \in [0, 1]\), followed by a final half oscillation. Let \(\xi_f\) denote the contribution from the full oscillations and \(\xi_h\) the contribution from the last half-cycle. We then have
\begin{align}
    \xi_f &= \int_0^{y\left(1 + \frac{m}{n} - \frac{1}{n}\right)} dy\, \cos\left(n\pi p(y)\right)\,, \\
    \label{defxih}
    \xi_h &= \int_{y\left(1 + \frac{m}{n} - \frac{1}{n}\right)}^{1 + \frac{m}{n}} dy\, \cos\left(n\pi p(y)\right)\,.
\end{align}
The contribution \(\xi_f\) spans an integer number of oscillations and can be evaluated using the same method as in the even \(n + m\) case. Applying Eq.(\ref{evenxi}), we obtain
\begin{align}
    \xi_f &= \frac{1}{n^2 \pi^2} \left( \frac{p''(0)}{p'(0)^3} - \frac{p''\left(y\left(1 + \frac{m}{n} - \frac{1}{n}\right)\right)}{p'\left(y\left(1 + \frac{m}{n} - \frac{1}{n}\right)\right)^3} \right) + \mathcal{O}\left( \frac{1}{n^3} \right) \\
    &= \frac{1}{n^2 \pi^2} \left( \frac{p''(0)}{p'(0)^3} - \frac{p''(1)}{p'(1)^3} \right) + \mathcal{O}\left( \frac{1}{n^3} \right)\,.
\end{align}
In the second step, we note that the effect of truncating the final half-cycle introduces an additional factor of \(1/n\), which can be absorbed into the \(\mathcal{O}(1/n^3)\) remainder.

We next evaluate the contribution from the final half oscillation, \(\xi_h\). Changing the integration variable to \(p\), we write
\begin{align}
    \xi_h &= \int_{1 + \frac{m}{n} - \frac{1}{n}}^{1 + \frac{m}{n}} dp\, \frac{dy}{dp} \cos(n\pi p)\,.
\end{align}
Again, Taylor expanding \(\frac{dy}{dp}\) around \(p = 1 + \frac{m}{n}\), we obtain
\begin{align}
    \xi_h &= \int_{1 + \frac{m}{n} - \frac{1}{n}}^{1 + \frac{m}{n}} dp \left[ \left( \frac{dy}{dp} \right)_{1 + \frac{m}{n}} + \left( \frac{d^2 y}{dp^2} \right)_{1 + \frac{m}{n}} \left(p - 1 - \frac{m}{n}\right) + \dots \right] \cos(n\pi p) \\
    &= -\left( \frac{d^2 y}{dp^2} \right)_{1 + \frac{m}{n}} \frac{1}{n^2 \pi^2} \int_{-\pi}^0 dq\, q \cos(q) + \mathcal{O}\left( \frac{1}{n^3} \right) \\
    &= -\left( \frac{d^2 y}{dp^2} \right)_{1 + \frac{m}{n}} \frac{2}{n^2 \pi^2} + \mathcal{O}\left( \frac{1}{n^3} \right) \\
    &= \frac{2}{n^2 \pi^2} \frac{p''(1)}{p'(1)^3} + \mathcal{O}\left( \frac{1}{n^3} \right)\,.
\end{align}
In the final step, we used Eq.(\ref{d2ydp2}) to express the result in terms of the function \(p(y)\) rather than its inverse. Combining \(\xi_f\) and \(\xi_h\), we obtain the total integral for odd \(n + m\):
\begin{align}
    \label{oddxi}
    \xi = \frac{1}{n^2 \pi^2} \left( \frac{p''(0)}{p'(0)^3} + \frac{p''(1)}{p'(1)^3} \right) + \mathcal{O}\left( \frac{1}{n^3} \right)\,.
\end{align}

\subsection{The Bogoliubov Coefficients \(\beta_{nm}\)}

Using Eq.(\ref{bln}) and the asymptotic results from Eq.(\ref{evenxi}) and Eq.(\ref{oddxi}), we can now write the large-\(n\) behavior of the Bogoliubov coefficients as
\begin{align}
    \beta_{nm} \approx -\sqrt{\frac{m}{n}} \frac{1}{n^2 \pi^2} \left( \frac{p''(0)}{p'(0)^3} - (-1)^{n + m} \frac{p''(1)}{p'(1)^3} \right)\,,
\end{align}
where the function \(p(y)\) is defined as
\begin{align}
    p(y) = \frac{\ln(1 + aL y)}{\ln(1 + aL)} + \frac{m}{n} y\,.
\end{align}
The sign alternation in this term arises from interference between oscillations reflected at the two boundaries of the box. Depending on whether the combined mode index $n+m$ is even or odd, these reflections interfere constructively or destructively, producing the alternating pattern in the excitation spectrum calculated later. This expression captures the leading-order behavior of \(\beta_{nm}\) in the regime \(n \gg 1\) for the general case of arbitrary acceleration, and will be instrumental in deriving the continuum excitation spectrum ahead.

\subsection{Excitation Spectrum}

\begin{figure}[h]
    \centering
    \includegraphics[width=0.7\linewidth]{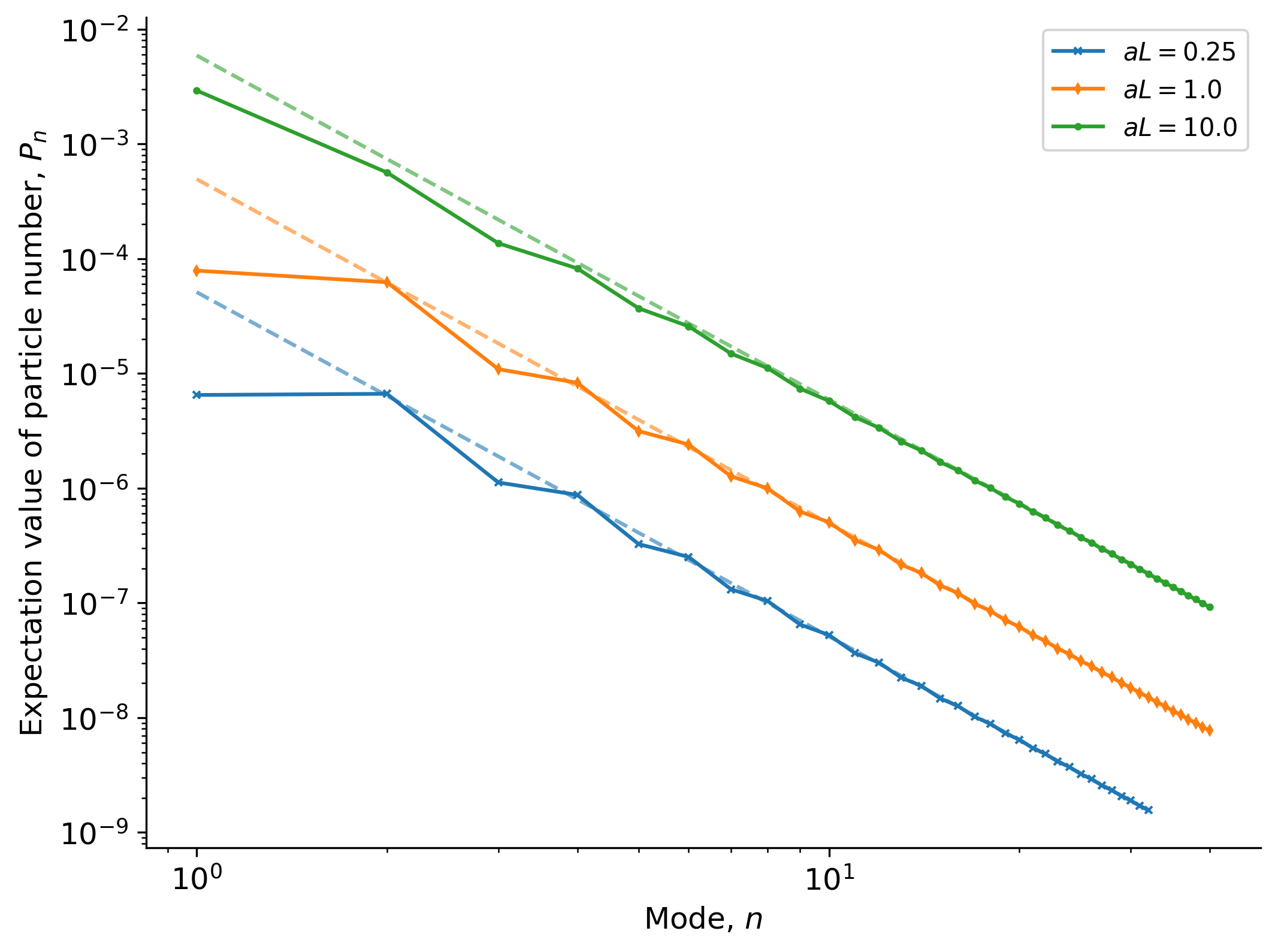}
    \caption{Exact spectra of excitations, as calculated before using Eq.(\ref{ab_exact}), shown on a log-log plot, with the large-\(n\) fit of Eq.(\ref{largen}) shown in dashed lines.}
    \label{fig:partfit}
\end{figure}

Having obtained the asymptotic form of \(\beta_{nm}\), we now compute the excitation spectrum \(P_n\), defined in Eq.(\ref{P}), which gives the expected number of particles created in the \(n^\text{th}\) mode. Separating contributions from even and odd values of \(n + m\), we write
\begin{equation}
    P_n = \left( \frac{1}{n^4 \pi^4} \sum_{\substack{m = 1 \\ n + m = \text{even}}}^{\infty} \left( \frac{p''(0)}{p'(0)^3} - \frac{p''(1)}{p'(1)^3} \right)^2 \right) + \left( \frac{1}{n^4 \pi^4} \sum_{\substack{m = 1 \\ n + m = \text{odd}}}^{\infty} \left( \frac{p''(0)}{p'(0)^3} + \frac{p''(1)}{p'(1)^3} \right)^2 \right)\,.
\end{equation}
In the large-\(n\) regime, we approximate the summations by integrals by introducing the continuous variable \(\zeta = m/n\). This yields
\begin{align}
    P_n &= \frac{1}{2 \pi^4 n^3} \left( \int_0^\infty d\zeta \left( \frac{p''(0,\zeta)}{p'(0,\zeta)^3} - \frac{p''(1,\zeta)}{p'(1,\zeta)^3} \right)^2 \right) + \frac{1}{2 \pi^4 n^3} \left( \int_0^\infty d\zeta \left( \frac{p''(0,\zeta)}{p'(0,\zeta)^3} + \frac{p''(1,\zeta)}{p'(1,\zeta)^3} \right)^2 \right) \\
    &= \frac{1}{\pi^4 n^3} \int_0^\infty d\zeta \left[ \left( \frac{p''(0,\zeta)}{p'(0,\zeta)^3} \right)^2 + \left( \frac{p''(1,\zeta)}{p'(1,\zeta)^3} \right)^2 \right]\,.
\end{align}
Here, we have made the \(\zeta\)-dependence of the function \(p(y)\) explicit:
\begin{align}
    p(y, \zeta) &= \frac{\ln(1 + aL y)}{\ln(1 + aL)} + \zeta y\,, \\
    p'(y, \zeta) &= \frac{aL}{\ln(1 + aL)} \cdot \frac{1}{1 + aL y} + \zeta\,, \\
    p''(y, \zeta) &= -\frac{a^2 L^2}{\ln(1 + aL)} \cdot \frac{1}{(1 + aL y)^2}\,.
\end{align}
Since \(p''(y, \zeta)\) is independent of \(\zeta\), the integrals simplify and resemble the form encountered in Eq.(\ref{smallint}). Evaluating the resulting expression, we obtain the asymptotic excitation spectrum:
\begin{align}
    \label{largen}
    P_n = \frac{1}{10 \pi^4} \cdot \frac{\left(\ln(1 + aL)\right)^2}{n^3}\,.
\end{align}
Figure~\ref{fig:partfit} shows the exact spectra \(P_n\) for various values of \(aL\), plotted on a log-log scale alongside the fit from Eq.(\ref{largen}). The agreement between the exact spectrum calculated numerically and the large \(n\) analytical expression is evident.

\subsection{Excitation Spectrum in Frequency Domain}

The expression in Eq.(\ref{largen}) is not only compact but also reveals deeper physical insights when interpreted in the frequency domain. In the large-\(n\) regime, the spacing between adjacent mode frequencies becomes small compared to the frequencies themselves. It is therefore natural to treat the frequency \(\Omega\) as a continuous variable and define a spectral density \(P(\Omega)\), such that \(P(\Omega)\, d\Omega\) gives the expected number of excitations in a small frequency interval of \(d\Omega\) around the frequency \(\Omega\).

An important point now is that observers located at different positions within the accelerating box will perceive different frequencies for the same mode due to gravitational\footnote{The uniform acceleration of the box is equivalent to a uniform gravitational field across the box. Hence, this time dilation can be interpreted as gravitational. Alternatively, one could just note that the Rindler metric is conformal, and the conformal factor varies with \(x\).} time dilation. The proper time \(\tau_O\) for an observer at proper distance \(\xi\) from the left edge of the box evolves as \(d\tau_O = (1 + a\xi)\, dt\). Using Eq.(\ref{freqleft}) for the frequency of the \(n^\text{th}\) mode as on the left edge, the frequency of the \(n^\text{th}\) mode as measured by the observer is
\begin{equation}
    \Omega_{n,O} = \frac{\Omega_n}{1 + a\xi} = \frac{n\pi a}{(1 + a\xi)\, \ln(1 + aL)}\,,
\end{equation}
and the spacing between adjacent frequencies is
\begin{equation}
    \Delta \Omega_O = \frac{\pi a}{(1 + a\xi)\, \ln(1 + aL)}\,.
\end{equation}

Rewriting Eq.(\ref{largen}) in terms of \(\Omega_n\) and dividing by \(\Delta \Omega_O\) to obtain the continuum spectrum, we find
\begin{equation}
    P(\Omega_O)\, d\Omega_O = \frac{1}{10 \pi^2} \cdot \frac{a^2}{(1 + a\xi)^2} \cdot \frac{1}{\Omega_O^3}\, d\Omega_O\,.
\end{equation}

Recognizing that \(a / (1 + a\xi)\) is indeed the proper acceleration \(a_O\) experienced by the observer, we arrive at a remarkably simple and universal expression:
\begin{equation}
    \label{PO}
    P(\Omega_O)\, d\Omega_O = \frac{1}{10 \pi^2} \cdot \frac{a_O^2}{\Omega_O^3}\, d\Omega_O\,.
\end{equation}

This result encapsulates several elegant features of the excitation spectrum.
\begin{enumerate}
    \item It follows an inverse cubic dependence on frequency.
    \item It is proportional to the square of the proper acceleration experienced by the observer.
    \item It is independent of the observer's position within the box.
    \item It is independent of the box's total length \(L\).
\end{enumerate}

These properties highlight the universality of the excitation spectrum in the frequency domain and hints at some deeper symmetry.

\section{Generalized Acceleration Profiles}

We now consider a broader class of scenarios beyond the sudden onset of uniform acceleration, motivated by the fact that realistic systems may experience time-dependent or stepwise changes in acceleration rather than a single instantaneous burst. Physically, such generalizations allow us to understand how the field responds dynamically to varying acceleration, and how particle production accumulates over multiple transitions.

\subsection{From \(a_i\) to \(a_f\)}

To start with, let's ask: what are the Bogoliubov coefficients when a box already undergoing uniform acceleration changes its acceleration at a given instant? Suppose the left edge of the box initially accelerates at rate \(a_i\), and this is changed to \(a_f\) at some instant, all the while as the box maintains constant proper length. What are the resulting Bogoliubov coefficients \(\alpha_{nm}\) and \(\beta_{nm}\) associated with this transition? What would be the spectrum of excitations created in the box, if the field was in the vacuum state before the transition (in the \(a_i\) phase of acceleration)?

Following the same procedure as before, that is, constructing mode expansions in the uniformly accelerated box before and after the change, and imposing continuity and differentiability conditions, we obtain a generalization of Eq.(\ref{I_m/n}) and Eq.(\ref{I_n/m}):
\begin{align}
    \begin{rcases}
        \alpha_{nm}(a_i \rightarrow a_f) \\
        \beta_{nm}(a_i \rightarrow a_f)
    \end{rcases}
    = \pm \sqrt{\frac{m}{n}} \frac{a_i}{1 + a_i L} \int_0^L \frac{dz}{1 + a_i z} \cos\left( n\pi \frac{\ln(1 + a_f z)}{\ln(1 + a_f L)} \mp \frac{\ln(1 + a_i z)}{\ln(1 + a_i L)} \right)
\end{align}
or equivalently,
\begin{align}
    \begin{rcases}
        \alpha_{nm}(a_i \rightarrow a_f) \\
        \beta_{nm}(a_i \rightarrow a_f)
    \end{rcases}
    = \sqrt{\frac{n}{m}} \frac{a_f}{1 + a_f L} \int_0^L \frac{dz}{1 + a_f z} \cos\left( n\pi \frac{\ln(1 + a_f z)}{\ln(1 + a_f L)} \mp \frac{\ln(1 + a_i z)}{\ln(1 + a_i L)} \right)\,.
\end{align}

Unlike Eq.(\ref{I_n/m}), these expressions do not admit closed-form solutions. However, they reveal a symmetry between the initial and final accelerations that was obscured in the earlier problem. Expanding to first order in \(a_i\) and \(a_f\), we obtain a generalization of Eq.(\ref{alphasmalla}) and Eq.(\ref{betasmalla}) in the small acceleration limit case as 
\begin{align}
    \alpha_{nm}^{(0)} &= \delta_{nm}\,, \\
    \beta_{nm}^{(0)} &= 0\,, \\
    \alpha_{nm}^{(1)} &= 
    \begin{cases}
        -(a_f - a_i) L \frac{2}{\pi^2} \frac{\sqrt{n m}}{(n - m)^3} & \text{if } n + m \text{ is odd} \\
        0 & \text{if } n + m \text{ is even}
    \end{cases} \\
    \label{smallaiafbeta}
    \beta_{nm}^{(1)} &= 
    \begin{cases}
        (a_f - a_i) L \frac{2}{\pi^2} \frac{\sqrt{n m}}{(n + m)^3} & \text{if } n + m \text{ is odd} \\
        0 & \text{if } n + m \text{ is even}
    \end{cases}
\end{align}

It is worth noting that for most realistic experimental setups, the dimensionless parameter \(aL / c^2\) is extremely small. Thus, Eq.(\ref{smallaiafbeta}) and the corresponding spectrum of excitations for a field initially in the vacuum state for large \(n\),
\begin{equation}
    \label{pngen}
    P_n \approx \frac{1}{10 \pi^4} \cdot \frac{(a_f - a_i)^2 L^2}{n^3}\,,
\end{equation}
are sufficient for practical analyses. Physically, the first-order \(\beta\)-coefficients, proportional to \(a_f - a_i\), indicate that particle creation arises from changes in acceleration rather than its absolute value; if the acceleration remains constant, no new particles are produced. Consequently, the excitation spectrum in Eq.~\eqref{pngen} scales with the square of the change in acceleration, reflecting the intensity of the perturbation imparted to the field.

\subsection{Successive acceleration steps}

Now consider a scenario where the box undergoes two successive changes in acceleration. Let the Bogoliubov coefficients for the first change be \(\alpha_{1,nm}\), \(\beta_{1,nm}\), and for the second change be \(\alpha_{2,nm}\), \(\beta_{2,nm}\). The cumulative Bogoliubov transformation from the initial to the final state is then given by
\begin{align}
    \alpha_{nm} &= \sum_{l = 1}^\infty \left( \alpha_{2,nl} \alpha_{1,lm} + \beta_{2,nl} \beta^*_{1,lm} \right)\,, \\
    \beta_{nm} &= \sum_{l = 1}^\infty \left( \alpha_{2,nl} \beta_{1,lm} + \beta_{2,nl} \alpha^*_{1,lm} \right)\,.
\end{align}
If both transitions involve small changes in acceleration, and we retain only first-order terms, we find
\begin{equation}
    \label{1ocomb}
    \beta_{nm}^{(1)} = \beta_{1,nm}^{(1)} + \beta_{2,nm}^{(1)}\,,
\end{equation}
i.e., the first-order Bogoliubov \(\beta\)-coefficients simply add over successive transitions.

To describe successive changes, we must introduce a time parameter, which leads to phase factors in the expression for \(\beta_{nm}\). Using Eq.(\ref{timefactor}), and retaining only first-order terms, we find that for a change in acceleration from \(a_i\) to \(a_f\) occurring at proper time \(\tau\),
\begin{equation}
    \label{fullsmallabeta}
    \beta_{nm} = \frac{(a_f - a_i) L}{\pi^2} \left(1 - (-1)^{n + m}\right) e^{i(n + m)\pi \frac{\tau}{L}} \frac{\sqrt{n m}}{(n + m)^3}\,.
\end{equation}
Combining expressions of the form Eq.(\ref{fullsmallabeta}) over multiple transitions using Eq.(\ref{1ocomb}), and generalizing to a continuously varying acceleration profile \(a(\tau)\), we obtain the elegant result:
\begin{equation}
    \label{beauty}
    \beta_{nm} = \frac{L}{\pi^2} \left(1 - (-1)^{n + m}\right) \frac{\sqrt{n m}}{(n + m)^3} \int d\tau\, \frac{da}{d\tau} e^{i(n + m)\pi \frac{\tau}{L}}\,.
\end{equation}
This expression captures the Bogoliubov coefficients for a box undergoing small, time-dependent accelerations, and highlights the linear response of the field to changes in acceleration. The exponential factor, \(\exp{i(n + m)\pi \tau / L}\), encodes the relative timing of the acceleration change: excitations produced at different times accumulate with different phases, leading to interference effects that can modulate the particle spectrum in scenarios with multiple transitions. Equation~\eqref{beauty} generalizes the discrete jumps to a continuously varying acceleration profile \(a(\tau)\). Further, the integral shows that the particle production is governed by the time derivative of the acceleration, \(da/d\tau\), confirming that the field responds dynamically to changes rather than absolute acceleration. This is analogous to a linear response in classical systems, where the response is proportional to the rate of change of the driving parameter. This expression agrees with an analogous expression found by Bruschi \emph{et al.}~\cite{Bruschi_2013_withoutparticlecreation} for the case where the box is inertial both at early and late times.

Finally, we note another maybe useful generalization: if the field is confined using Neumann boundary conditions, that is, requiring \(\partial \phi / \partial x = 0\) at the boundaries, instead of Dirichlet conditions, the results derived above remain unchanged, indicating that the qualitative features of particle creation, namely, its dependence on changes in acceleration, the linear superposition over multiple transitions, and the phase-sensitive interference are robust and largely independent of the precise boundary type. While we do not show this explicitly, it is straightforward to verify by starting from the modified mode functions with cosines instead of sines. The zeroth mode decouples from the system, and doesn't have or affect particle creation. For the rest of the modes, expressions analogous to Eq.(\ref{I}) and Eq.(\ref{J}) are found, with cosines replacing the sines. They can be manipulated by integrating by parts and combining to get the same expressions for \(\alpha_{nm}\) and \(\beta_{nm}\).

\section{Constraints and Challenges for Detection}

\begin{figure}[h]
    \centering
    \includegraphics[width=0.7\linewidth]{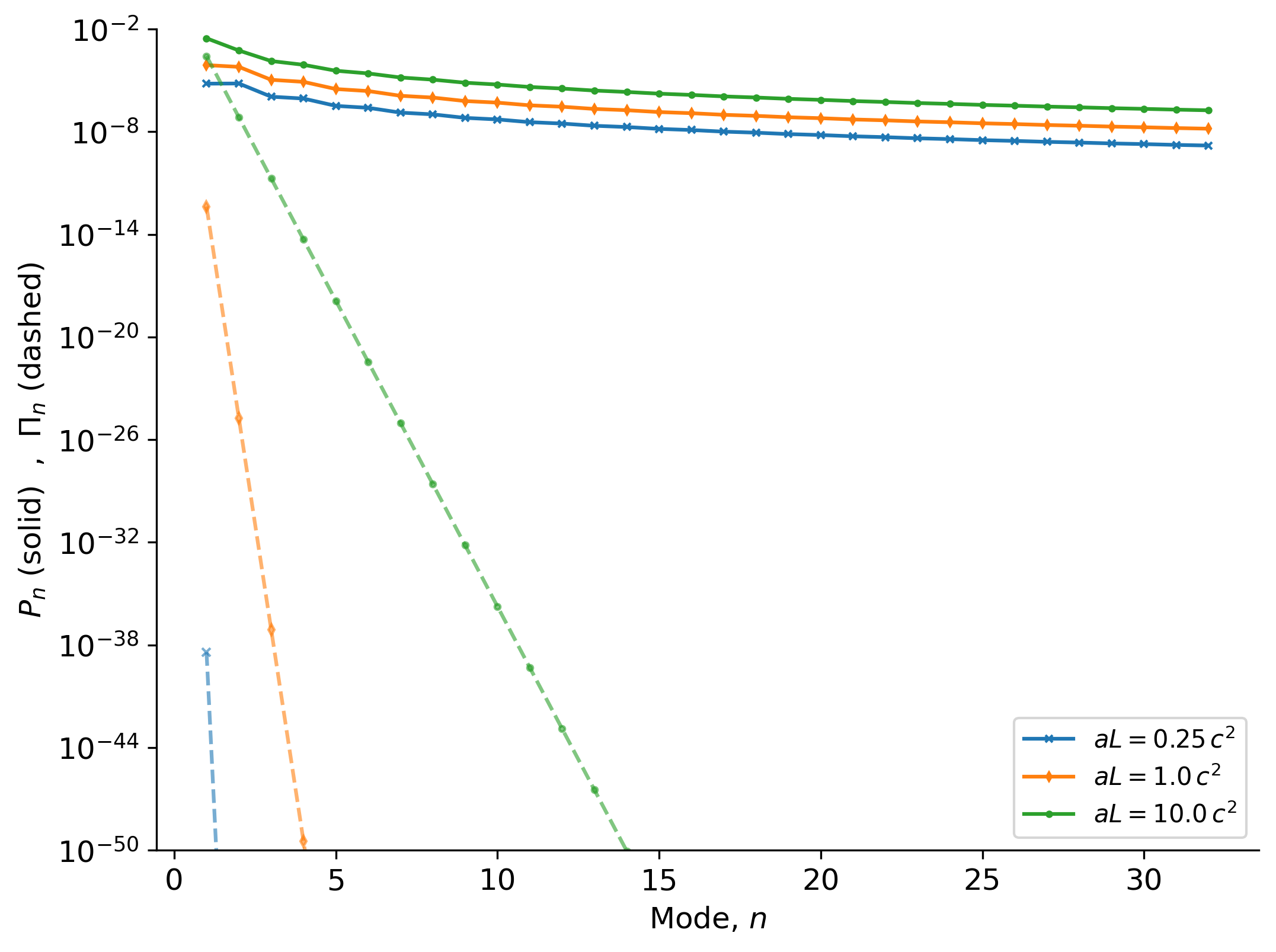}
    \caption{The particle spectrum for selected values of \(aL\), compared with the spectrum expected from an Unruh-like effect.}
    \label{fig:unruh}
\end{figure}

We begin by noting that the excitation spectrum \(P_n\), as given by Eq.(\ref{largen}), is orders of magnitude larger than what one would expect from an Unruh-type radiation effect. By an Unruh-type effect, we refer to a thermal spectrum of the form with a temperature given by the Unruh temperature, \(T_\text{Unruh} = a / 2\pi\). Using Eq.(\ref{freqleft}) for the frequency of the modes, this thermal spectrum would look like
\begin{equation}
    \Pi_n = \frac{1}{e^{\frac{n\pi a}{\ln(1 + aL)} \cdot \frac{2\pi}{a}} - 1}\,.
\end{equation}
A comparison between \(P_n\) and \(\Pi_n\) for various values of acceleration is shown in Fig.~\ref{fig:unruh}. For the small accelerations achievable in laboratory settings, \(aL\lesssim c^2\) and \(P_n\) clearly dominates and thus presents a more promising candidate for probing non-inertial quantum field effects.

Thermal noise can also be effectively suppressed, as its contribution decays exponentially with increasing mode number \(n\). The ratio of \(P_n\) to the thermal noise \(\mathcal{N}_n\) in the \(n^\text{th}\) mode at temperature \(T\) is given, using Eq.(\ref{pngen}), by
\begin{align}
    \frac{P_n}{\mathcal{N}_n} = \frac{\frac{1}{10\pi^4} \cdot \frac{(a_f - a_i)^2 L^2}{n^3}}{\frac{1}{e^{\frac{n\pi}{L T}} - 1}} = \left( e^{\frac{n\pi \hbar c}{L k_B T}} - 1 \right) \cdot \frac{1}{10\pi^4} \cdot \frac{(a_f - a_i)^2 L^2}{c^4 n^3}\,,
\end{align}
where we have restored physical units in the final expression. A modest reduction in temperature exponentially enhances the signal-to-noise ratio \(P_n / \mathcal{N}_n\), making low-temperature rigid boxes or cavities a viable platform for detection.

\begin{figure}[h]
    \centering
    \includegraphics[width=0.7\linewidth]{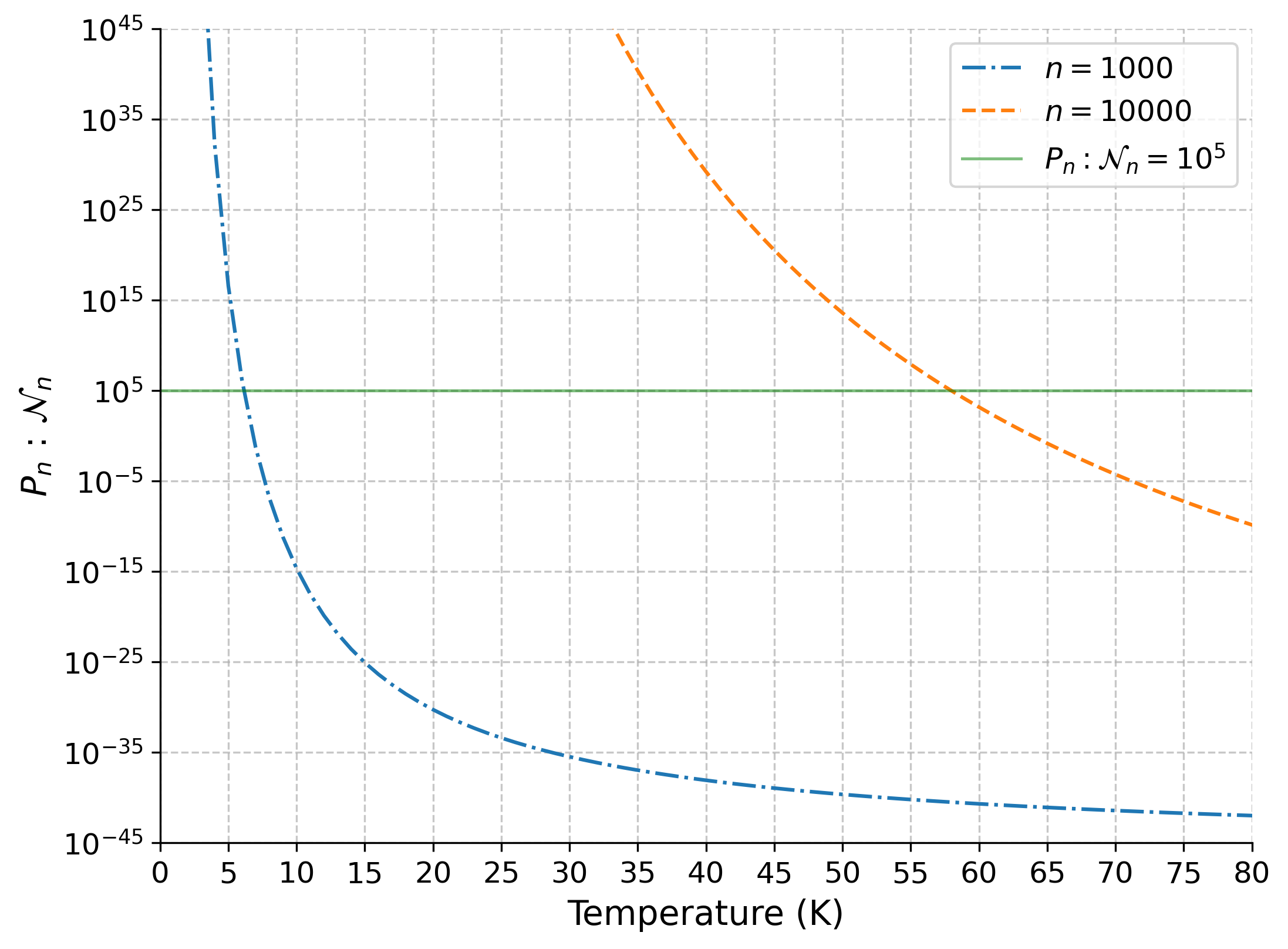}
    \caption{Variation of the ratio of particle number produced due to acceleration to thermal noise with temperature.}
    \label{fig:snr}
\end{figure}

As an example, consider a perfectly reflecting metal cavity of width \(L = 0.01\, \text{m}\), undergoing a change in acceleration of \(100\, \text{m/s}^2\) (approximately \(10g\)).\footnote{While the electromagnetic field is not a scalar field, the mode functions for a photon is the same as for a free massless scalar field, with an added complication of polarization. We will not worry about this, and use the results above as it is for photons in a reflecting cavity.} Suppose we observe photons in the modes \(n = 1000\) and \(n = 10000\)\footnote{Both of these modes would lie roughly in the infrared regime}. Figure~\ref{fig:snr} shows the variation of the signal to noise ratio, \(P_n / \mathcal{N}_n\), with temperature for these modes. To detect photons, sufficiently clearly above the (thermal) noise, say near \(n = 10000\), the cavity must be cooled to approximately \(60\, \text{K}\), while for \(n = 1000\), a temperature of around \(5\, \text{K}\) is required, both of which are well within reach of current cryogenic technology.

Another consideration is the detector's\footnote{Here, by a detector, we mean any mechanism inserted into the cavity to detect the field. We do not consider the Unruh de-Witt detector formalism to rigorously note the response of a detector.} influence on the mode structure. To avoid significant perturbation, the detector must interact weakly with the field. If placed at the cavity walls, it should detect only a small fraction \(f\) of the reflected photons; if placed within the cavity, it should intercept only a fraction \(f\) of the passing photons. To ensure all photons are eventually detected, the cavity must maintain its acceleration profile for a duration sufficient for photons to traverse the cavity approximately \(1/f\) times. This sets a lower bound on the experiment duration of
\begin{equation}
    \delta t \sim \frac{2L}{f c}\,.
\end{equation}
For the parameters discussed above and \(f = 0.001\), we find
\begin{equation}
    \label{deltat}
    \delta t \approx 7 \times 10^{-8}\, \text{sec}\,,
\end{equation}
a duration modestly achievable in practice.

The most significant challenge lies in the smallness of \(P_n\) itself. For meaningful detection, we would ideally require \(P_n \sim 1\), which is far from the case. A naive strategy would be to repeat the experiment many times and use a broadband detector. However, this quickly proves impractical. Suppose we employ a detector with bandwidth \(\mu n\), centered at mode \(n\). The total number of particles detected is
\begin{equation}
    P_\text{det} = \sum_{n' = n - \frac{\mu}{2} n}^{n + \frac{\mu}{2} n} P_{n'}\,.
\end{equation}
Using Eq.(\ref{pngen}) and approximating the sum by an integral, we obtain
\begin{equation}
    P_\text{det} = \frac{1}{10\pi^4} \cdot \frac{(a_f - a_i)^2 L^2}{n^2} \int_{1 - \frac{\mu}{2}}^{1 + \frac{\mu}{2}} dx\, \frac{1}{x^3} = \frac{8}{5\pi^4} \cdot \frac{\mu}{(4 - \mu^2)^2} \cdot \frac{(a_f - a_i)^2 L^2}{n^2}\,.
\end{equation}
For a reasonable bandwidth \(\mu = 1\), i.e., detecting modes between \(n/2\) and \(3n/2\), we find
\begin{equation}
    P_\text{det} = \frac{8}{45\pi^4} \cdot \frac{(a_f - a_i)^2 L^2}{c^4 n^2}\,,
\end{equation}
where we have restored dimensions. For the setup described earlier, with \(a_f - a_i = 100\, \text{m/s}^2\) and \(L = 0.01\, \text{m}\), we obtain
\begin{equation}
    P_\text{det} \approx 2 \times 10^{-43}
\end{equation}
for \(n = 1000\), and a hundredfold smaller value for \(n = 10000\). To detect even a single particle, the experiment would need to be repeated \(10^{43}\) times. Even if we change acceleration every \(10^{-7}\) seconds, the fastest permissible given Eq.(\ref{deltat}), the total duration required would be
\[
10^{36}\, \text{sec} \sim 3 \times 10^{28}\, \text{years}\,,
\]
rendering such a detection infeasible with current technology.

Now, the extra spatial dimensions we inhabit offer a potential advantage in amplifying the excitation spectrum. While a full three-dimensional calculation is considerably more involved, we can perform a simple order-of-magnitude estimate to assess the impact.

Let the cavity have dimensions \((L, L_Y, L_Z)\), where \(L\) is aligned with the direction of acceleration, and \(L_Y\), \(L_Z\) span the transverse directions. The field modes are now labeled by a triplet \((n, n_Y, n_Z)\), corresponding to longitudinal and transverse wavenumbers. For small \(aL\), and for modes satisfying
\begin{equation}
    \label{smallparap}
    \frac{n_Y}{L_Y}, \frac{n_Z}{L_Z} \lesssim \frac{n}{L}\,,
\end{equation}
i.e., transverse wavenumbers significantly smaller than the longitudinal wavenumber, the mode functions in the inertial and accelerated boxes can be approximated as
\begin{align}
    \phi_\text{in}(T,X) &= \sum_{n=1}^{\infty} \frac{1}{\sqrt{n\pi}} \sin\left(n\pi \frac{X}{L}\right) \sin\left(n_Y\pi \frac{Y}{L_Y}\right) \sin\left(n_Z\pi \frac{Z}{L_Z}\right) \left( b_n e^{-i n\pi \frac{T}{L}} + b_n^\dagger e^{i n\pi \frac{T}{L}} \right)\,, \\
    \phi_\text{ac}(t,x) &= \sum_{n=1}^{\infty} \frac{1}{\sqrt{n\pi}} \sin\left(n\pi \frac{x}{l}\right) \sin\left(n_Y\pi \frac{Y}{L_Y}\right) \sin\left(n_Z\pi \frac{Z}{L_Z}\right) \left( c_n e^{-i n\pi \frac{t}{l}} + c_n^\dagger e^{i n\pi \frac{t}{l}} \right)\,.
\end{align}
The key approximation here is
\begin{equation}
    \sqrt{\left( \frac{n\pi}{L} \right)^2 + \left( \frac{n_Y\pi}{L_Y} \right)^2 + \left( \frac{n_Z\pi}{L_Z} \right)^2} \approx \frac{n\pi}{L}\,,
\end{equation}
which is justified under the condition in Eq.(\ref{smallparap}). Additionally, we approximate the accelerated mode functions using sinusoidal profiles as in the inertial case, which is why it is only valid for small \(aL\). Under these assumptions, continuity and differentiability of the field across the transition yield a Bogoliubov transformation of the form
\begin{align}
    \alpha^\text{3D}_{(n,n_Y,n_Z)(m,m_Y,m_Z)} &= \alpha_{nm} \delta_{n_Y m_Y} \delta_{n_Z m_Z}\,, \\
    \beta^\text{3D}_{(n,n_Y,n_Z)(m,m_Y,m_Z)} &= \beta_{nm} \delta_{n_Y m_Y} \delta_{n_Z m_Z}\,,
\end{align}
where \(\alpha_{nm}\) and \(\beta_{nm}\) are the one-dimensional Bogoliubov coefficients previously derived.

Consequently, the excitation spectrum for each mode \((n, n_Y, n_Z)\) is simply
\begin{equation}
    P^\text{3D}_{(n,n_Y,n_Z)} = P_n\,,
\end{equation}
where \(P_n\) is the one-dimensional excitation spectrum. The total number of excitations in the \(n^\text{th}\) longitudinal mode is obtained by summing over all transverse modes satisfying Eq.(\ref{smallparap}). The number of such modes scales as
\[
\sim \left( \frac{L_Y}{L} n \right) \left( \frac{L_Z}{L} n \right) = n^2 \frac{L_Y L_Z}{L^2}\,,
\]
yielding
\begin{equation}
    P^\text{3D}_n \sim P_n \cdot n^2 \frac{L_Y L_Z}{L^2}\,.
\end{equation}
We make no assumptions about the contribution from high transverse wavenumbers; if significant, they would only enhance \(P^\text{3D}_n\). Substituting \(P_n\) from Eq.(\ref{pngen}), we obtain
\begin{equation}
    P^\text{3D}_n \sim \frac{1}{10\pi^4} \cdot \frac{(a_f - a_i)^2 L_Y L_Z}{c^4 n}\,.
\end{equation}
If a detector is sensitive to photons with transverse wavenumbers between \(n_\text{min}\) and \(n_\text{max}\), the total number of photons detected is
\begin{equation}
    P^\text{3D}_\text{det} \sim \frac{1}{10\pi^4} \cdot \frac{(a_f - a_i)^2 L_Y L_Z}{c^4} \cdot \ln\left( \frac{n_\text{max}}{n_\text{min}} \right)\,.
\end{equation}
Since \(\ln(n_\text{max}/n_\text{min}) \sim 1\) for any reasonable bandwidth, and \(10\pi^4 \sim 1000\), we estimate, for the previous configuration with \(a_f - a_i = 100\, \text{m/s}^2\) and \(L_Y = L_Z = 10\, \text{m}\),
\begin{equation}
    P^\text{3D}_\text{det} \sim \frac{(a_f - a_i)^2 L_Y L_Z}{1000 c^4} \approx 10^{-28}\,.
\end{equation}
Assuming acceleration changes every \(10^{-7}\, \text{s}\), we would require \(10^{21}\, \text{s} \sim 3 \times 10^{13}\) years to detect a single particle—an improvement over previous estimates, though still far from feasible. If we consider extreme acceleration values \(\sim 400{,}000\, g\)\footnote{This is the acceleration that can be achieved in a modern centrifuge}, we obtain
\[
P^\text{3D}_\text{det} \sim 2 \times 10^{-22}\,,
\]
which would require approximately 20 million years for detection under the same repetition rate.

An alternative direction for experimental feasibility is to explore the possibility of rapidly and precisely modulating the acceleration of the cavity with frequencies comparable to the frequency of the mode we wish to excite. In the small \(aL\) regime, Eq.(\ref{beauty}) provides a compact expression for the Bogoliubov coefficients \(\beta_{nm}\), where the integral involves \(\dot{a}(\tau)\) multiplied by a rapidly oscillating exponential. Suppose we subject the cavity to a time-dependent acceleration of the form
\[
a(\tau) = 2a_0 \cos\left( \frac{k\pi \tau}{L} \right)\,, \quad k \in \mathbb{Z}^+\,,
\]
i.e., an oscillating acceleration at the frequency of a standing mode inside the box. Substituting into Eq.(\ref{beauty}), we obtain
\begin{align}
    \beta_{nm} &= \frac{L}{\pi^2} \left(1 - (-1)^{n + m} \right) \frac{\sqrt{nm}}{(n + m)^3} \cdot i \frac{k\pi}{L} \int d\tau\, \left( a_0 e^{i k\pi \frac{\tau}{L}} + a_0 e^{-i k\pi \frac{\tau}{L}} \right) e^{i(n + m)\pi \frac{\tau}{L}} \\
    &\approx \frac{a_0 L}{\pi^2} \left(1 - (-1)^{n + m} \right) \frac{\sqrt{nm}}{(n + m)^3} \cdot i \frac{k\pi}{L} \cdot \delta_{k, n + m} \cdot T\,,
\end{align}
where in the second line we have assumed a long interaction time \(T\), allowing the integral to be approximated by a delta function in frequency space. Rewriting this expression with dimensions restored, we find
\begin{equation}
    \beta_{nm} \approx i \frac{a_0 T}{\pi c} \left(1 - (-1)^k \right) \frac{\sqrt{n(k - n)}}{k^2} \delta_{m, k - n}\,.
\end{equation}

This result reveals several important features: (i) Constructive excitation occurs only for odd values of \(k\), due to the factor \(1 - (-1)^k\). (ii) For a given \(k\), only modes with \(n < k\) contribute, and for each such \(n\), there is a unique \(m = k - n\) that satisfies the resonance condition. (iii) The excitation spectrum \(P_n = \sum_m |\beta_{nm}|^2\) is hence just \(|\beta_{n,n-k}|^2\), giving
\begin{equation}
    \label{pnparabola}
    P_n = 4 \frac{a_0^2 T^2}{\pi^2 c^2} \cdot \frac{n(k - n)}{k^4}\,,
\end{equation}
which describes an inverted parabola peaking at \(n \approx k/2\). To target excitations around \(n \sim 1000\), we require \(k \sim 2000\), say \(k = 2001\) (odd). For a cavity of length \(L \approx 0.01\, \text{m}\), this corresponds to an acceleration oscillation frequency of approximately \(30\, \text{THz}\), with a precision better than \(0.05\%\). While challenging, let's assume for now that such control is achievable.

Next, we consider the contribution from transverse modes. For a given longitudinal mode \(n\), the number of transverse modes contributing with spectrum given by Eq.(\ref{pnparabola}) is approximately
\[
\sim \left( \frac{L_Y}{L} \right) \left( \frac{L_Z}{L} \right) n^2\,.
\]
The total number of excitations produced is then, for odd \(k\),
\begin{align}
    P_\text{tot} &\sim 4 \frac{a_0^2 T^2}{\pi^2 c^2} \cdot \frac{L_Y L_Z}{L^2} \sum_{n = 1}^{k} \frac{n(k - n)}{k^2} n^2 \\
    &\approx 4 \frac{a_0^2 T^2}{\pi^2 c^2} \cdot \frac{L_Y L_Z}{L^2} \cdot k \int_0^1 d\mu\, \mu^3 (1 - \mu) \\
    &= \frac{a_0^2 T^2}{5 \pi^2 c^2} \cdot \frac{L_Y L_Z}{L^2} \cdot k\,.
\end{align}
Assuming lateral dimensions \(L_Y = L_Z = 1\, \text{m}\), \(k = 2000\), and peak acceleration \(a_0 = 15000\, g\)\footnote{This is the typical rating for electronics in artillery applications, so, we know that we have electrical systems that can survive these accelerations.}, we find that it takes approximately \(10\, \text{seconds}\) for \(P_n \sim 1\). Moreover, since \(P_\text{tot} \propto T^2\), extending the duration of oscillation leads to a quadratic increase in particle production.

These are interesting estimates. However, unlike the case of a single acceleration change, where detection can occur continuously, here we must preserve the quantum state of the field and ensure that it is undisturbed for the entire duration \(T\). This requires near-perfect confinement of the field and minimal coupling to the external environment. While technically demanding, such isolation is increasingly feasible given the rapid advances in quantum technologies, particularly those driven by quantum computing and precision control of quantum systems.

\section{Discussion}

In this work, we have investigated the excitation of scalar field modes confined within a rigid cavity undergoing sudden or time-dependent acceleration. Starting from exact expressions for the Bogoliubov coefficients in terms of incomplete gamma functions, we obtained the excitation spectrum \(P_n\) and analyzed its behavior across different regimes. In the low-acceleration limit, compact analytic forms of \(\alpha_{nm}\) and \(\beta_{nm}\) were derived, exhibiting an alternating pattern in mode excitations and a universal decay proportional to \(n^{-3}\). This scaling was corroborated through an exact large-\(n\) analysis, yielding
\[
\frac{1}{10 \pi^4} \cdot \frac{\left(\ln(1 + aL)\right)^2}{n^3}.
\]
The universality of the \(n^{-3}\) decay reflects the robustness of acceleration-induced particle creation, independent of microscopic details of the box.

Recasting the results in the frequency domain, we found that the spectral density \(P(\Omega_O)\) measured by any comoving detector depends solely on the proper acceleration \(a_O\) and scales as \(a_O^2 / \Omega_O^3\). Remarkably, this dependence is independent of the observer’s position or the cavity’s proper size. For general, time-dependent acceleration profiles, the Bogoliubov coefficients acquire a convolution form, from which we showed that resonant, oscillatory accelerations at odd cavity-mode frequencies lead to selective amplification of excitations. Such resonant configurations offer promising routes for experimental verification in optomechanical or superconducting-circuit platforms.

The asymptotic form of the excitation spectrum, \(P_n \sim a^2/n^3\), bears a close analogy to the Larmor formula for classical radiation, where the emitted power scales as \(a^2\). In both classical and quantum settings, acceleration serves as a source of radiation, where the former produces electromagnetic waves, the latter generates quanta from vacuum fluctuations. This correspondence, previously noted in single-mirror setups \cite{NikishovRitus1995}, reinforces the conceptual continuity between classical and quantum radiation mechanisms \cite{Higuchi:1992we, Higuchi:1992td, Landulfo:2019tqj, Paithankar:2020akh}.

Equally striking is the universality of the frequency-domain spectrum. That \(P(\Omega_O)\) depends only on the observer’s acceleration, and not on their spatial position or the cavity dimensions, hints at a connection to horizon physics. In the limit where the cavity is extended indefinitely, while effectively forming a Rindler wedge, the dominant feature is the causal restriction: radiation within the wedge cannot communicate with the exterior, a significant divergence from the structure underlying the Unruh effect.

The present analysis also clarifies a subtle aspect relevant to experimental searches for acceleration-induced radiation. Any realistic attempt to compare inertial and accelerated observers necessarily involves a finite-time transition between frames. Such transient effects modify the instantaneous particle content, leading to excitations that may far exceed the weak thermal signal associated with the Unruh temperature \(T = a/2\pi\). In principle, these transients could thus provide a more accessible observational signature than the steady-state Unruh radiation itself.

Finally, the mechanism described here may have astrophysical relevance. In extreme environments, such as magnetized plasma sheets or accretion flows, localized regions can experience rapid, non-uniform accelerations while maintaining partial confinement. If such systems act analogously to rigid cavities, they could produce measurable signatures of acceleration-induced quantum excitations. Although speculative, these possibilities underscore the broader reach of quantum field theory in non-inertial settings.

\section*{Acknowledgments}
The authors thank Jorma Louko for valuable discussions and insightful comments on this manuscript.

\bibliographystyle{unsrt}
\bibliography{references}
\end{document}